\documentclass[conference]{IEEEtran}
\IEEEoverridecommandlockouts

\usepackage{amsmath,amssymb,amsfonts}
\usepackage{comment}
\usepackage{cite}
\usepackage{framed}
\usepackage{color}
\usepackage{listings}
\usepackage{latexsym}
\usepackage{graphicx}
\usepackage{url}

\usepackage{algpseudocode}

\usepackage[linesnumbered,vlined,boxed,commentsnumbered]{algorithm2e}

\usepackage{comment}
\usepackage{setspace}
\usepackage[T1]{fontenc}

\begin{document}

\title{Generating Shuttling Procedures for Constrained Silicon Quantum Dot Array}

\author{\IEEEauthorblockN{Naoto Sato, Tomonori Sekiguchi, Takeru Utsugi, and Hiroyuki Mizuno}
\IEEEauthorblockA{\textit{Research \& Development Group,} 
\textit{Hitachi, Ltd.}\\
naoto.sato.je@hitachi.com}

}

\maketitle

\begin{abstract}
In silicon quantum computers, a single electron is trapped in a microstructure called a quantum dot, and its spin is used as a qubit. For large-scale integration of qubits, we previously proposed an approach of arranging the quantum dots in a two-dimensional array and sharing a control gate in a row or column of the array. In our array, the shuttling of electrons is a useful technique to operate the target qubit independently and avoid crosstalk. However, since the shuttling is also conducted using shared control gates, the movement of qubits is complexly constrained. We therefore propose a formal model on the basis of state transition systems to describe those constraints and operation procedures on the array. We also present an approach to generate operation procedures under the constraints. Utilizing this approach, we present a concrete method for our 16 $\times$ 8 quantum dot array. By implementing the proposed method as a quantum compiler, we confirmed that it is possible to generate operation procedures in a practical amount of time for arbitrary quantum circuits. We also demonstrated that crosstalk can be avoided by shuttling and that the fidelity in that case is higher than when crosstalk is not avoided.
\end{abstract}

\section{Introduction}\label{intro}
Silicon quantum computing based on mature semiconductor technology is a promising approach for the large-scale integration of qubits. A single electron is trapped in a microstructure called a quantum dot formed in silicon substrate, and its spin is used as a qubit. In order to execute a quantum gate operation on those qubits, it is necessary to connect control gates (plungers/barriers) to each quantum dot. As the number of qubits increases, the number of control gates and their wirings also increases. Since sufficient space is required to implement a large number of qubits, this imposes a limitation for large-scale integration. Our approach to solving this problem is to arrange the quantum dots in a two-dimensional array and provide a shared control gate connected to the quantum dots in the same row or column of the array \cite{Lee_2020}\cite{Lee_2022}.

There are three challenges when it comes to implementing quantum gate operations in the quantum dot array using our approach. 
The first challenge is how to connect a qubit arranged regularly in a two-dimensional array to read-out circuits that can be placed on the rows at the end of the array. 
The second challenge is how to suppress of the crosstalk during the quantum gate operation. In a single qubit operation (e.g., spin rotation to a qubit), the Larmor frequency of the qubit is selectively matched with the frequency of an externally applied RF wave. However, neighboring qubits around the target qubit also receive the effects of the RF wave from this operation, known as crosstalk, and its spin direction is slightly disturbed. This reduces the fidelity of the neighboring qubits.
The third challenge is how to ensure the independent quantum gate operation of a single qubit in an array where control gates are shared by multiple quantum dots in a column or row. For example, when a quantum gate operation is executed on a qubit using a control gate shared by the column, another qubit in that column will also be operated on. 

To overcome these challenges, we have proposed {\it shuttling qubits} \cite{mizuno2023quantum} \cite{sekiguchiSiQEW}. An electron in a quantum dot can be shuttled to a neighboring quantum dot if there is no electron in the dot, which is executed by applying appropriate control voltage to the control gates. Regarding the first challenge, an arbitrary qubit can be shuttled to the read-out circuit in the end rows of the array. Crosstalk, the second challenge, can be suppressed by evacuating the neighboring qubits away from the target qubit before the quantum gate operation. For the third challenge, independent operation can be achieved by moving other qubits away from the target qubit. To allow horizontal and vertical shuttling of qubits, we proposed two-dimensional control gates \cite{Lee_2020}. The control gates in the first layer are shared by quantum dots in the same column of the array. A qubit is shuttled horizontally using the column-shared control gates in the first layer. If certain conditions are met, only one qubit can be shuttled independently. Otherwise, all qubits in the same column move simultaneously in the same direction. The shuttling in vertical directions is conducted by the row-shared control gates in the second layer. A row-shared control gate is connected to several quantum dots in the row, and the qubits in the connected quantum dots are vertically shuttled by the row-shared control gate simultaneously in the same direction.
These complex constraints of the quantum dot array make it difficult for the quantum compiler to find a procedure for operating the quantum dot array in a practical amount of time to execute arbitrary sequences of quantum gates, that is, quantum circuits.

The first contribution of this paper is to present a formal model of quantum dot arrays for expressing the operation procedures and constraints of the array. We then demonstrate the usefulness of the model by formally describing the structural and operational constraints of our 16 $\times$ 8 quantum dot array \cite{Lee_2020}\cite{Lee_2022} and the constraints to avoid crosstalk. The second contribution is to show an approach to generate operation procedures under those complex constraints. On the basis of this approach, we propose a concrete method to generate operation procedures for our 16 $\times$ 8 quantum dot array as the third contribution. By implementing the proposed method as a quantum compiler, we confirmed that it is possible to generate operation procedures in a practical amount of time for arbitrary quantum circuits. Furthermore, simulations confirmed that crosstalk can be avoided in the generated operation procedures, and that the procedure that avoids crosstalk by shuttling has higher fidelity of the output quantum state than the procedure that does not avoid crosstalk.

In Section \ref{prelim} of this paper, we briefly introduce quantum circuits and silicon quantum dot arrays. The constraints and requirements that the operation procedures of our silicon quantum dot array should satisfy are formally defined in Section \ref{constraint}. The problem we tackle in this paper is described with examples in Section \ref{problem}. As an approach to solve the problem, we present a state transition system and its conditions to search for the operation procedure in Section \ref{secapp}. In Section \ref{instance}, we present a concrete state transition system for our 16 $\times$ 8 silicon quantum dot array. The algorithm to search the transition system and extract an operation procedure is shown in Section \ref{algorithm}. The proposed method is implemented and evaluated in Section \ref{exp}. We discuss the improvements of the transition system to obtain more efficient operation procedures and the effectiveness of crosstalk avoidance in Section \ref{discuss}. In Section, \ref{relwork} related work is described. The conclusions drawn from this study are presented in Section \ref{conclusion}.

\section{Preliminaries}\label{prelim}

\subsection{Quantum Circuit}
\begin{figure}[htb]
\centering
\scalebox{0.5}{\includegraphics{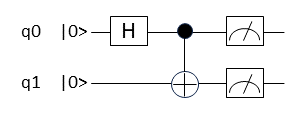}}
\caption{Quantum circuit to generate the Bell state}
\label{bell}
\end{figure}

A qubit, which is a variable in a quantum program, is in a superposition of the basis state |0> and |1>. When a qubit is measured, 0 or 1 is observed probabilistically, depending on its superposition state. The state of a qubit can be expressed as $|\psi> = a_0 |0> + a_1 |1>$ ($|a_0|^2 + |a_1|^2=1$), where $a_0$ and $ a_1$ are complex numbers and called amplitudes. The absolute squares of the amplitudes $|a_0|^2$ and $|a_1|^2$ represent the probabilities of obtaining 0 and 1 by a measurement. Generally, an arbitrary quantum state consisting of $n$ qubits is represented by $2^n$ basis states.
A quantum program is represented by a model called a quantum circuit. Figure \ref{bell} shows an example of a quantum circuit to generate the Bell state. Each horizontal line corresponds to a qubit, and the operations on them, called quantum gates, are arranged from left to right. 
In this quantum circuit, qubits $q0$ and $q1$ are both initialized to $|0>$, respectively. Then, after the Hadamard gate is executed for $q0$, the CNOT gate is executed on $q0$ and $q1$. Both $q0$ and $q1$ are then measured.

\subsection{Silicon Quantum Dot Array}\label{array}
\begin{figure}[htb]
\centering
\scalebox{0.5}{\includegraphics{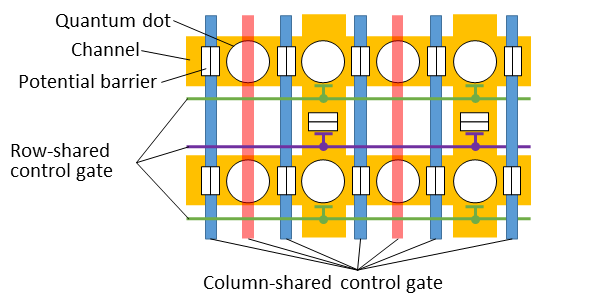}}
\caption{Structure of quantum dot array}
\label{quantumdotarray}
\end{figure}

In our silicon quantum dot array (SQDA), quantum gate operations and measurements are conducted by control gates that are connected to quantum dots. If one control gate is connected to each quantum dot, the number of control gates increases with the number of qubits. Therefore, reducing the number of control gates is a problem that needs to be solved for large-scale integration of qubits. To this end, we proposed arranging the quantum dots in a two-dimensional array and sharing a control gate in each row and column.
The structure of the quantum dot array presented in our prior work \cite{Lee_2020}\cite{Lee_2022} is shown in Fig. \ref{quantumdotarray}.

The black circles represent quantum dots, and the thick orange lines are the channels connecting the quantum dots. Each quantum dot is controlled by a control gate connected to that quantum dot and its adjacent potential barrier. The rectangles represent potential barriers between quantum dots. The red and green lines are the column-shared and row-shared control gates for the quantum dots (plunger gates), respectively. Similarly, the blue and purple lines are column-shared control gates for the potential barrier between horizontally adjacent quantum dots and row-shared control gates for the potential barrier between vertically adjacent quantum dots (barrier gates), respectively.

When control gates are shared for each row and column of the array, the number of control gates increases by $O(\sqrt{n})$ for the number of qubits $n$. This is advantageous for large-scale integration compared to the case where control gates are connected individually to each qubit. However, another problem arises: the same operation is always conducted on quantum dots connected to the same control gate. This problem can be solved by shuttling electrons between quantum dots \cite{Spiderweb}\cite{Mills2019}\cite{BlueprintofaScalableSpin}\cite{Acrossbarnetwork}\cite{Simulatedcoherentelectron}\cite{ShuttlinganElectronSpin}\cite{Fujita2017}\cite{Nakajima2018}\cite{Yoneda2021}\cite{Noiri2022}. Before executing an operation on a particular dot, electrons in quantum dots connected to the same control gate as the dot can be evacuated by shuttling. This allows the operation on a specific qubit. In addition, silicon qubits can suffer from interference, called crosstalk \cite{philips2022universal}\cite{heinz2022high}\cite{heinz2021crosstalk}\cite{heinz2022crosstalk}\cite{hsiao2020efficient}\cite{throckmorton2022crosstalk}, between neighboring qubits during qubit gate operations, resulting in reduced fidelity. By evacuating irrelevant electrons from around the qubits to be operated on in advance, the occurrence of crosstalk can be avoided and the reduction in fidelity can be prevented. 
Thus, by utilizing electron shuttling, it may be possible to reduce both the number of control gates and the crosstalk. It should be noted, however, that electron shuttling is also conducted by the same control gates as quantum gate operations. In other words, the simplification of the quantum dot array structure to reduce the number of control gates also imposes restrictions on the shuttling of electrons. This shuttling constraint is formulated in the next section.

The native quantum gate set of the SQDA consists of $Rx(\theta)$ and $Ry(\theta)$ gates for arbitrary $ \theta $ and a $(SWAP)^{\alpha}$ gate where the exponent $ \alpha $ is arbitrarily controlled by adjusting the strength and duration of the Heisenberg exchange \cite{nielsen2010quantum}\cite{fan2005optimal}.
An arbitrary quantum circuit can be decomposed into an equivalent quantum circuit consisting of these native gates. We assume that the quantum circuit consists of only these native gates. The difference between the control of the $Rx(\theta)$ and $Ry(\theta)$ gates is the phase of the irradiated radio frequency electromagnetic wave, and the control related to the electron transfer is the same. Therefore, in the following, we refer to the $Rx(\theta)$ and $Ry(\theta)$ gates together as a single-qubit gate. Similarly, the $(SWAP)^{\alpha}$ gate is referred to as a two-qubit gate. Quantum circuits that appear in the following are assumed to consist of only these native gates. 

\section{Constraints and Requirements of Array Operation Procedure}\label{constraint}
As mentioned in Section \ref{array}, in the SQDA shown in our prior work \cite{Lee_2020}\cite{Lee_2022}, control gates are shared for each column and row. This has the constraint that the same quantum gate operations, measurements, and shuttling operations are executed on electrons located in the same column or row at the same time. In this section, we formally define these SQDA constraints as properties that array operation procedures should satisfy. Similarly, using the same formal model, we define the requirement that the quantum gate operations be executed as indicated by the quantum circuit as a property of the array operation procedure.

\subsection{Formal Model of SQDA}\label{fmodel}
The structure of a quantum dot array is represented as an undirected graph. The movement of electrons on the array by shuttling is modeled as a state transition system $M$, where the state represents the position of the electrons on the graph. $M$ is denoted by $(S, s_0, L, T, F)$. $S$ is a finite set of states and $s_0$ is an initial state ($s_0 \in S$). $L$ is a finite set of labels that correspond to the operations of quantum gate, measurement, and shuttling. $T$ is a set of transitions defined as $T \subseteq (S \times L \times S)$. $F$ is a set of final states. Since operations on the SQDA can be stopped at any time, $F$ includes all states in $S$, that is, $F = S$. A finite transition path $p$ of $M$ is given as a sequence of transitions $ [ t_{1}, ..., t_{i}, ..., t_{n} ] $. Transition $t_i \in T$ is expressed as $(s_{i-1}, l_{i}, s_{i})$, where $s_{i-1}, s_{i} \in S$ and $l_{i} \in L$. 

An arbitrary state $s_i$ is denoted by $(A, R, B, pos_i)$. The $A$ represents the topology of the array, which is invariant at any $s_i$ since the topology does not change at runtime. $A$ is represented by a graph $(V, E)$ with quantum dots as vertices and the channels connecting them as edges. The set of vertices $V$ is the set of quantum dots. The set of edges $E$ is the set of channels, whose elements are unordered pairs of two quantum dots connected by channels. That is, $E \subseteq \{ (x,y) |x,y \in V \land x \neq y \}$. In our SQDA, quantum dots are arranged in a lattice. The number of rows and columns of the array is described as $ROW$ and $COL$, respectively. We denote the quantum dots in row $r \leq ROW$ and column $c \leq COL$ as $V_{r, c}$. $V$ is defined using $v_{r, c}$, as in formula \ref{fdot}.
\begin{align}
V {\mathrm{def}}{=} \left\{ v_{r,c} | 1 \leq r \leq ROW \land 1 \leq c \leq COL \right\} \label{fdot}
\end{align}

Channels are formed between particular adjacent quantum dots. That is, the following formula \ref{fchannel} holds for $E$.
\begin{align}
\forall v_{r1,c1}, v_{r2,c2} \cdot (v_{r1,c1}, v_{r2,c2}) \in E \nonumber \\
\Rightarrow ( |r1-r2|+|c1-c2|=1 ) \label{fchannel}
\end{align}

\begin{figure}[htb]
\centering
\scalebox{0.5}{\includegraphics{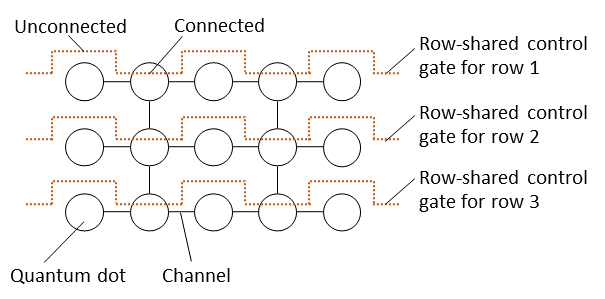}}
\caption{Example of row-shared control gates}
\label{2gvertical}
\end{figure}

$R : V \rightarrow Bool$ is a predicate indicating that the quantum dots are connected to row-shared control gates. Between quantum dots that are vertically connected in a channel, a two-qubit gate is executed using row-shared control gates. The row-shared control gate is not connected to all the qubits in the row. However, if a quantum dot is connected to a row-shared control gate, all the quantum dots in the same column are connected to the row-shared control gate located in the respective row. This constraint is represented by formula \ref{frwire}.
\begin{align}
\forall r1, r2, c \cdot 1 \leq r1 \leq ROW \land 1 \leq r2 \leq ROW \nonumber \\
\land 1 \leq c \leq COL \land R(v_{r1,c}) \Rightarrow R(v_{r2,c}) \label{frwire}
\end{align}
The structure of quantum dots, channels, and row-shared control gates is illustrated in Fig. \ref{2gvertical}. For the sake of simplicity, the column-shared control gates are not shown because they are installed in all rows. More precisely, as shown in Fig. \ref{quantumdotarray}, some column-shared control gates (red line) are not connected to the quantum dots that are connected to the row-shared control gate. However, the same control as that by the column control gate can be achieved by the row-shared control gate (green line). Therefore, we can assume that all quantum dots are connected to the column-shared control gate, and the column-shared control gates are thus omitted in Fig. \ref{2gvertical} and subsequent figures.

The $B$ is the set of electrons on the array, and an electron is described as $b_j \in B$. Function $pos_i$ represents the position of the electrons at $s_i$. It maps each electron to the quantum dot in which it exists, i.e., $pos_i: B \rightarrow V$. For use in the following formulae, we define the set of quantum dots $P_i \subseteq V$ as follows: $P_i = \left\{ v | b \in B \land v = pos_i(b) \right\}$. $P_i$ represents the quantum dots that trap an electron in $s_i$.

The label $l_i \in L$ is represented by a set of tuples consisting of electrons and an operation on them. There are four types of operations: single-qubit gate, two-qubit gate, measurement, and shuttling. The operator for the single-qubit gate operation is denoted by $g1$, and the single-qubit gate operation on electron $b_j$ is denoted by $(g1, b_j)$. Similarly, a measurement on $b_j$ is denoted by $(m, b_j)$. The two-qubit gate operation on electrons $b_{j1}$ and $b_{j2}$ is denoted as $(g2, b_{j1}, b_{j2})$. Since electrons may move up, down, left, or right, we define $sh \text{-}u$, $sh \text{-} d$, $sh \text{-} l$, and $sh \text{-} r$ as shuttling operators, respectively. For example, the operation to shuttle electron $b_j$ in the upward direction is denoted as $(sh \text{-} u, b_j)$. Label $l_i$ is the set of these operations, that is, multiple operations can be executed simultaneously in a single state transition.
Note that the states in $M$ represents not quantum states but the positions of electrons on the SQDA. The operations of quantum gates and measurements in which the electron positions do not change are represented as loop transitions from a state to the same state. That is, if the label $l_i$ is a set consisting of arbitrary quantum gate and measurement operations, $s_{i-1} = s_i$ at the transition $(s_{i-1}, l_i, s_i)$.

\subsection{Constraints of Quantum Gates and Measurements}\label{gateconst}
Before defining the constraints, we introduce the predicates $SameCol_i$, $SameRow_i$, $AdjHor_i$, and $AdjVer_i$ as predicates related to the position of electrons on the array. The predicate $SameCol_i(b_j, b_k)$ will be True if $b_j$ and $b_k$ are in the same column in state $s_i$. Similarly, $SameRow_i(b_j, b_k)$ will be True if $b_j$ and $b_k$ are in the same row in state $s_i$. The predicates $AdjHor_i(b_j, b_k)$ will be True if $b_{j}, b_{k}$ are horizontally adjacent and the quantum dots that trap them are connected by channels. $AdjHor_i$ means that these two electrons can be executed on by a two-qubit gate. Similarly, the predicates $AdjVer_i(b_j, b_k)$ will be True if $b_{j}, b_{k}$ are vertically adjacent and the quantum dots that trap them are connected by channels. In addition, it should be held that these quantum dots have to be connected to row-shared control gates because the vertical two-qubit gate is not conducted by column-shared gates but by row-shared control gates. The formal definitions of these predicates are given in Appendix \ref{predicatedef}.

\subsubsection{Single-qubit Gates}
\begin{figure}[htb]
\centering
\scalebox{0.4}{\includegraphics{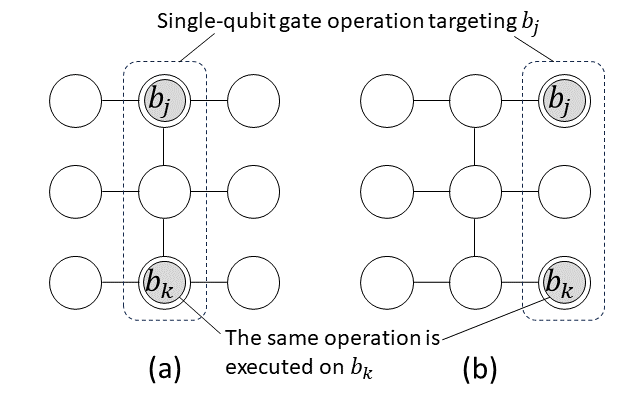}}
\caption{Single-qubit gate operation}
\label{1qubitoperation}
\end{figure}

A single-qubit gate operation is conducted using a column-shared control gate. Therefore, when a single-qubit gate operation is executed on an electron $b_j$, the same single-qubit gate operation is also executed on the other electrons in the same column, such as $b_k$ in Fig. \ref{1qubitoperation}(a)(b). 
That is, formula \ref{f1qubit} is satisfied, where $l_{i+1}$ is the label of arbitrary transition from state $s_i$.
\begin{align}
\forall b_j, b_k \in B, \forall l_{i+1} \in L \cdot (g1, b_j) \in l_{i+1} \nonumber \\
\land SameCol_i(b_j, b_k) \Rightarrow (g1, b_k) \in l_{i+1} \label{f1qubit}
\end{align}

\begin{figure}[htb]
\centering
\scalebox{0.4}{\includegraphics{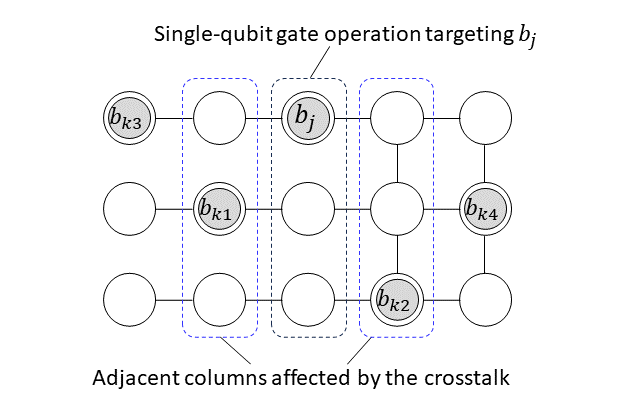}}
\caption{Crosstalk of single-qubit gate operation}
\label{1qubitcross}
\end{figure}

Since single-qubit gates are executed by column-shared control gates, all electrons in adjacent columns to the right and left of the target electrons are affected by crosstalk. For example, in Fig. \ref{1qubitcross}, electrons $b_{k1}$ and $b_{k2}$ are affected by crosstalk of the operation targeting electron $b_j$. Electrons $b_{k3}$ and $b_{k4}$ are not affected. 
Therefore, when executing a single-qubit gate, it is necessary to ensure that there is no electron in the adjacent quantum dot in the controlled column. This constraint is expressed by the following formula \ref{f1gatecross}.
\begin{align}
\forall b_j \in B, \forall l_{i+1} \in L, \forall v_{r_k,c_k} \in V \cdot 
(g1, b_j) \in l_{i+1} \nonumber \\
\land v_{r_j,c_j} = pos_i(b_j)
\land |c_j - c_k| = 1 
\Rightarrow 
v_{r_k,c_k} \notin P_i \label{f1gatecross}
\end{align}

\subsubsection{Two-qubit Gates}
\begin{figure}[htb]
\centering
\scalebox{0.4}{\includegraphics{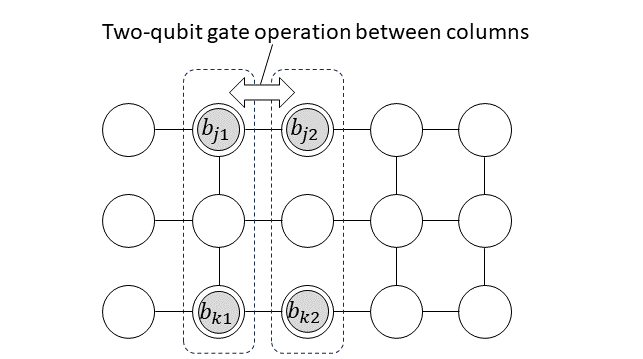}}
\caption{Horizontal two-qubit gate operation}
\label{2qubithor}
\end{figure}

The two-qubit gate operation can be executed between two adjacent quantum dots connected by a channel horizontally or vertically. This is expressed by the following formula \ref{f2qubit_channel}.
\begin{align}
\forall b_{j1}, b_{j2} \in B, \forall l_{i+1} \in L \cdot (g2, b_{j1}, b_{j2}) \in l_{i+1} \nonumber \\
\Rightarrow \left( AdjHor_i(b_{j1}, b_{j2}) \lor AdjVer_i(b_{j1}, b_{j2}) \right) \label{f2qubit_channel}
\end{align}
When the target electrons $b_{j1}$ and $b_{j2}$ are horizontally adjacent, column-shared control gates are used to execute a two-qubit gate. Therefore, the two-qubit gate operation is also executed on the other two electrons $b_{k1}$ and $b_{k2}$ that are in the same columns and are horizontally adjacent (Fig. \ref{2qubithor}). 
The following formula \ref{f2qubithor} expresses this constraint.
\begin{align}
& \forall b_{j1}, b_{j2}, b_{k1}, b_{k2} \in B, \forall l_{i+1} \in L \cdot (g2, b_{j1}, b_{j2}) \in l_{i+1} \nonumber \\
& \land AdjHor_i(b_{j1}, b_{j2}) \land AdjHor_i(b_{k1}, b_{k2}) \nonumber \\
& \land SameCol_i(b_{j1}, b_{k1}) \land SameCol_i(b_{j2}, b_{k2}) \nonumber \\
& \Rightarrow (g2, b_{k1}, b_{k2}) \in l_{i+1} \label{f2qubithor}
\end{align}

\begin{figure}[htb]
\centering
\scalebox{0.4}{\includegraphics{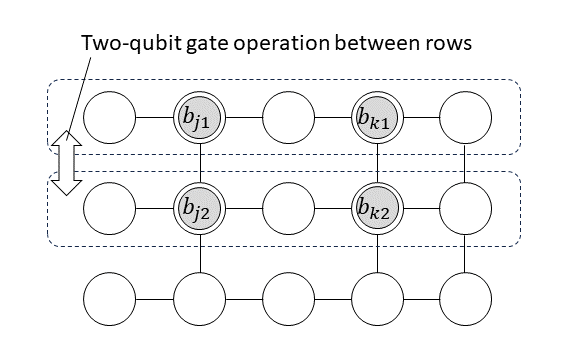}}
\caption{Vertical two-qubit gate operation}
\label{2qubitver}
\end{figure}

When the target electrons $b_{j1}$ and $b_{j2}$ are vertically adjacent, row-shared control gates are used. Therefore, the same two-qubit gate is executed on the other two electrons $b_{k1}$ and $b_{k2}$ in the same rows and are vertically adjacent (Fig. \ref{2qubitver}). 
This constraint can be expressed by the following formula \ref{f2qubitver}.
\begin{align}
& \forall b_{j1}, b_{j2}, b_{k1}, b_{k2} \in B, \forall l_{i+1} \in L \cdot (g2, b_{j1}, b_{j2}) \in l_{i+1}  \nonumber \\
& \land AdjVer_i(b_{j1}, b_{j2}) \land AdjVer_i(b_{k1}, b_{k2}) \nonumber \\
& \land SameRow_i(b_{j1}, b_{k1}) \land SameRow_i(b_{j2}, b_{k2}) \nonumber \\
& \Rightarrow (g2, b_{k1}, b_{k2}) \in l_{i+1} \label{f2qubitver}
\end{align}

\begin{figure}[htb]
\centering
\scalebox{0.4}{\includegraphics{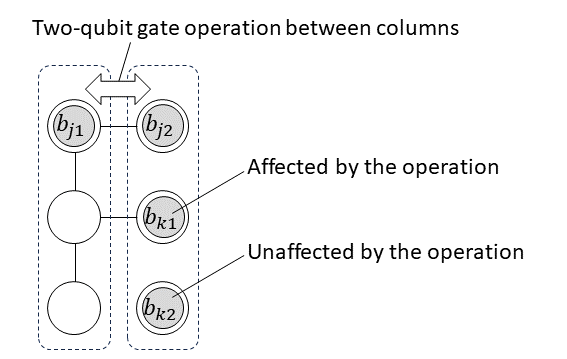}}
\caption{Effects of two-qubit gate operation}
\label{2qubitcross}
\end{figure}

In the case of a two-qubit gate, i.e., $(SWAP)^{\alpha}$, it can be assumed that no crosstalk occurs in the adjacent quantum dots \cite{martins2016noise}\cite{reed2016reduced}.
However, if another electron exists in a quantum dot in the target columns or rows in which the two-qubit gate operation is executed, that electron will be affected by the two-qubit gate operation. Therefore, it is necessary to evacuate the other electrons from the target columns or rows before executing the two-qubit gate. Note that if the quantum dot in which an electron exists is not connected to the adjacent quantum dot by a channel, that electron is not affected by the two-qubit gate operation. In Fig. \ref{2qubitcross}, electron $b_{k1}$ is affected by the two-qubit gate operation, but $b_{k2}$ is not affected.
In the case of horizontal two-qubit gates, the constraint to avoid the effect is expressed by the following formula \ref{fcross2qubithor}.
\begin{align}
& \forall b_{j1}, b_{j2} \in B, \forall l_{i+1} \in L, \forall v_{r_{k},c_{k1}}, v_{r_{k},c_{k2}} \in V \cdot  \nonumber \\
& (g2, b_{j1}, b_{j2}) \in l_{i+1} \land AdjHor_i(b_{j1}, b_{j2}) \nonumber \\ 
& \land v_{r_{j},c_{j1}} = pos_i(b_{j1}) \land v_{r_{j},c_{j2}} = pos_i(b_{j2}) \nonumber \\
& \land c_{k1} = c_{j1} \land c_{k2} = c_{j2} \land (v_{r_{k},c_{k1}}, v_{r_{k},c_{k2}}) \in E \nonumber \\
& \Rightarrow ( v_{r_{k},c_{k1}} \notin P_i \land v_{r_{k},c_{k2}} \notin P_i ) \nonumber \\
& \lor ( v_{r_{k},c_{k1}} \in P_i \land v_{r_{k},c_{k2}} \in P_i ) \label{fcross2qubithor}
\end{align}
For vertical two-qubit gates, the constraint is described as formula \ref{fcross2qubitver}.
\begin{align}
& \forall b_{j1}, b_{j2} \in B, \forall l_{i+1} \in L, \forall v_{r_{k1},c_{k}}, v_{r_{k2},c_{k}} \in V \cdot \nonumber \\ 
& (g2, b_{j1}, b_{j2}) \in l_{i+1} \land AdjVer_i(b_{j1}, b_{j2}) \nonumber \\ 
& \land v_{r_{j1},c_{j}} = pos_i(b_{j1}) \land v_{r_{j2},c_{j}} = pos_i(b_{j2}) \nonumber \\
& \land r_{k1} = r_{j1} \land r_{k2} = r_{j2} \land (v_{r_{k1},c_{k}}, v_{r_{k2},c_{k}}) \in E \nonumber \\
& \Rightarrow ( v_{r_{k1},c_{k}} \notin P_i \land v_{r_{k2},c_{k}} \notin P_i ) \nonumber \\
& \lor ( v_{r_{k1},c_{k}} \in P_i \land v_{r_{k2},c_{k}} \in P_i ) \label{fcross2qubitver}
\end{align}

\subsubsection{Measurements}
\begin{figure}[htb]
\centering
\scalebox{0.4}{\includegraphics{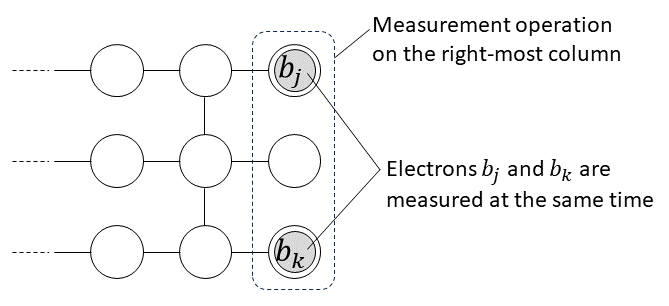}}
\caption{Measurement operation}
\label{measureoperation}
\end{figure}

In our SQDA, a measurement operation can be conducted on the rightmost column of the array. If there are other electrons in the rightmost column, the same measurement operation is executed on the other electrons, as well as the single-qubit gate (Fig. \ref{measureoperation}). 
These constraints are represented by the following formulae \ref{fmeasure1} and \ref{fmeasure2}, where $COL$ represents the rightmost column number of the array.
\begin{align}
\forall b_j \in B, \forall l_{i+1} \in L \cdot (m, b_{j}) \in l_{i+1} \nonumber \\
\Rightarrow v_{r_j,c_j} = pos_i(b_j) \land c_j = COL \label{fmeasure1}
\end{align}
\begin{align}
\forall b_j, b_k \in B, \forall l_{i+1} \in L \cdot (m, b_j) \in l_{i+1} \nonumber \\
\land SameCol(b_j, b_k) \Rightarrow (m, b_k) \in l_{i+1} \label{fmeasure2}
\end{align}

\subsection{Shuttling of Electrons}
An electron in one quantum dot can be shuttled only to the adjacent quantum dot to which the channel is connected. If an electron is moved to a quantum dot where another electron is already trapped, the electrons collide and quantum information is lost, so shuttling is prohibited in such cases. Since shuttling in the vertical direction is conducted by row-shared control gates, the quantum dot in which the target electron exists has to be connected to row-shared control gates. These constraints are expressed by the following formulae \ref{fshconst1} to \ref{fshconst4}.
\begin{align}
& \forall b_j \in B, \forall l_{i+1} \in L \cdot (sh \text{-} u, b_j) \in l_{i+1} \nonumber \\
& \Rightarrow v_{r_j,c_j} = pos_i(b_j) \land r_j > 1 \land R(v_{r_j,c_j})\nonumber \\
& \land \left( v_{r_j,c_j}, v_{r_{j}-1,c_j} \right) \in E \land v_{r_{j}-1,c_j} \notin P_i \label{fshconst1}
\end{align}
\begin{align}
& \forall b_j \in B, \forall l_{i+1} \in L \cdot (sh \text{-} d, b_j) \in l_{i+1} \nonumber \\
& \Rightarrow v_{r_j,c_j} = pos_i(b_j) \land r_j < ROW \land R(v_{r_j,c_j}) \nonumber \\
& \land \left( v_{r_j,c_j}, v_{r_{j}+1,c_j} \right) \in E \land v_{r_{j}+1,c_j} \notin P_i \label{fshconst2}
\end{align}
\begin{align}
& \forall b_j \in B, \forall l_{i+1} \in L \cdot (sh \text{-} l, b_j) \in l_{i+1} \nonumber \\
& \Rightarrow v_{r_j,c_j} = pos_i(b_j) \land c_j > 1 \nonumber \\
& \land \left( v_{r_j,c_j}, v_{r_{j},c_{j}-1} \right) \in E \land v_{r_{j},c_{j}-1} \notin P_i \label{fshconst3}
\end{align}
\begin{align}
& \forall b_j \in B, \forall l_{i+1} \in L \cdot (sh \text{-} r, b_j) \in l_{i+1} \nonumber \\
& \Rightarrow v_{r_j,c_j} = pos_i(b_j) \land c_j < COL \nonumber \\
& \land \left( v_{r_j,c_j}, v_{r_{j},c_{j}+1} \right) \in E \land v_{r_{j},c_{j}+1} \notin P_i \label{fshconst4}
\end{align}

\begin{figure}[htb]
\centering
\scalebox{0.5}{\includegraphics{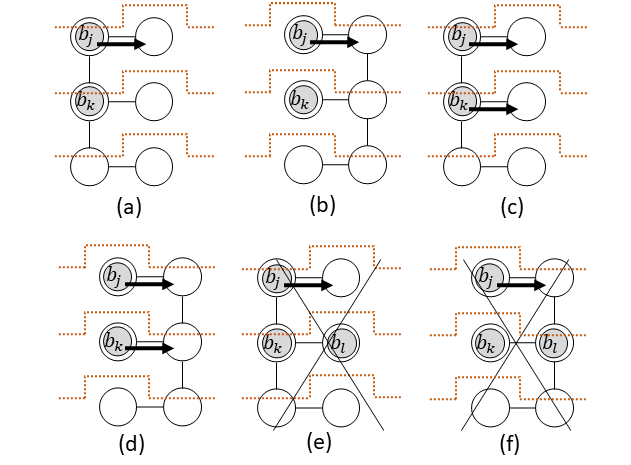}}
\caption{Horizontal shuttling with block control}
\label{blockcontrol}
\end{figure}

\begin{figure*}[htb]
\centering
\scalebox{0.5}{\includegraphics{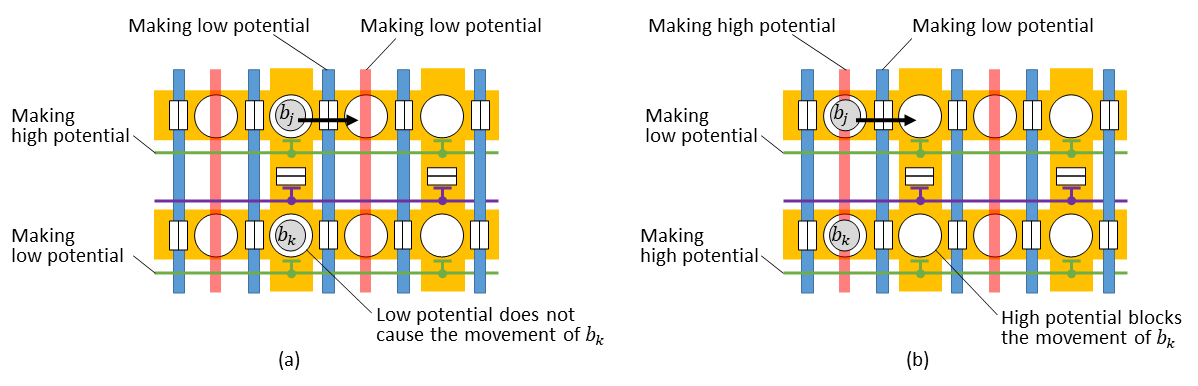}}
\caption{Potential control for block control}
\label{array_blockcontrol}
\end{figure*}

The shuttle operation in the horizontal direction is conducted by controlling the column-shared control gates. Therefore, when an electron is shuttled horizontally, other electrons in the same column are also shuttled in the same direction. However, it is possible to block the shuttling of other electrons in the same column, which is called {\it block control} \cite{utsugi2023single}.
The block control is implemented by raising the potential barrier between the quantum dots. This barrier prevents other electrons from moving at the same time as the target electron.
The conditions under which the shuttling of electrons in the same column can be blocked are shown in Fig. \ref{blockcontrol}.
When $b_j$ is shuttled, electron $b_k$ in the same column can be blocked if the quantum dot in the column with $b_j$ is connected to a row-shared control gate and the quantum dot in the column to which $b_j$ is moving is not connected to a row-shared control gate (Fig. \ref{blockcontrol}(a)). This is achieved by changing the voltage applied to the row-shared control gates of $b_j$ and $b_k$. A high potential is formed in the row of $b_j$ and a low potential in the row of $b_k$ (Fig. \ref{array_blockcontrol}(a)). This makes a potential difference between the current quantum dot of $b_j$ and its right quantum dot, and only $b_j$ is shuttled to the right. Since there is not a potential difference between the current quantum dot of $b_k$ and its right quantum dot, $b_k$ is not moved. Similarly, $b_k$ can be blocked if the quantum dot in the column with $b_j$ is not connected to a row-shared control gate and the quantum dot in the column to which $b_j$ is moving is connected to a row-shared control gate (Fig. \ref{blockcontrol}(b)). This is also achieved by controlling the row-shared control gates of $b_j$ and $b_k$ separately. A low potential is formed in the row of $b_j$ and a high potential in the row of $b_k$ (Fig. \ref{array_blockcontrol}(b)). The high potential prevents $b_k$ from shuttling to the right. 
Note that in both cases, it is also possible to move $b_k$ at the same time as $b_j$ (Fig. \ref{blockcontrol}(c)(d)). However, if there is another electron in the quantum dot to which $b_k$ is moving, $b_k$ cannot be blocked. In that case, to avoid the collision of the electrons, the shuttling operation on $b_j$ is prohibited (Fig. \ref{blockcontrol}(e)(f)). Even if the above conditions for block control are not satisfied, $b_k$ is blocked if there is no channel between the quantum dot of $b_k$ and the quantum dot to which $b_k$ is moving. These constraints of shuttling are expressed by the following formulae \ref{fblockctrl1} and \ref{fblockctrl2}. The predicate $BC$ is defined by the formula \ref{fbccondition} in Appendix \ref{predicatedef}, which will be True if the conditions for block control are satisfied.
\begin{align}
& \forall b_j, b_k \in B, \forall l_{i+1} \in L \cdot (sh \text{-} l, b_j) \in l_{i+1} \nonumber \\
& \land SameCol_i(b_j, b_k) \land v_{r_k,c_k} = pos_i(b_k) \nonumber \\
& \land \lnot BC(v_{r_k,c_k}, v_{r_{k},c_{k}-1}) \land ( v_{r_k,c_k}, v_{r_{k},c_{k}-1} ) \in E \nonumber \\
& \Rightarrow (sh \text{-} l, b_k) \in l_{i+1} \label{fblockctrl1}
\end{align}
\begin{align}
& \forall b_j, b_k \in B, \forall l_{i+1} \in L \cdot (sh \text{-} r, b_j) \in l_{i+1} \nonumber \\
& \land SameCol_i(b_j, b_k) \land v_{r_k,c_k} = pos_i(b_k) \nonumber \\
& \land \lnot BC(v_{r_k,c_k}, v_{r_{k},c_{k}+1}) \land ( v_{r_k,c_k}, v_{r_{k},c_{k}+1} ) \in E \nonumber \\
& \Rightarrow (sh \text{-} r, b_k) \in l_{i+1} \label{fblockctrl2}
\end{align}

\begin{figure}[htb]
\centering
\scalebox{0.5}{\includegraphics{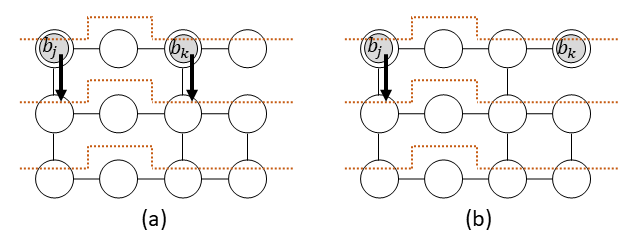}}
\caption{Vertical shuttling}
\label{shuttlever}
\end{figure}

The vertical shuttling operation is conducted by the row-shared control gates. Therefore, if a quantum dot that is in the same row as the target electron $b_j$ and is connected to the row-shared control gate traps the electron $b_k$, $b_k$ is shuttled together in the same vertical direction as $b_j$ (Fig. \ref{shuttlever}(a)). If there is no channel between the quantum dot of $b_k$ and the quantum dot to which it shuttles, $b_k$ stays (Fig. \ref{shuttlever}(b)). 
The constraints on the vertical shuttling are expressed by formulae \ref{fblockctrl3} and \ref{fblockctrl4}. Note that in the case of vertical shuttling, $b_k$ cannot be blocked because the potential barrier cannot be controlled separately for each column.
\begin{align}
& \forall b_j, b_k \in B, \forall l_{i+1} \in L \cdot (sh \text{-} u, b_j) \in l_{i+1} \nonumber \\
& \land SameRow_i(b_j, b_k) \land v_{r_k,c_k} = pos_i(b_k) \land R(v_{r_k,c_k}) \nonumber \\
& \land ( v_{r_k,c_k}, v_{r_{k}+1,c_k}) \in E \Rightarrow (sh \text{-} u, b_k) \in l_{i+1} \label{fblockctrl3}
\end{align}
\begin{align}
& \forall b_j, b_k \in B, \forall l_{i+1} \in L \cdot (sh \text{-} d, b_j) \in l_{i+1} \nonumber \\
& \land SameRow_i(b_j, b_k) \land v_{r_k,c_k} = pos_i(b_k) \land R(v_{r_k,c_k}) \nonumber \\
& \land ( v_{r_k,c_k}, v_{r_{k}-1,c_k}) \in E \Rightarrow (sh \text{-} d, b_k) \in l_{i+1} \label{fblockctrl4}
\end{align}

In addition, we define a constraint about the simultaneous execution of single-qubit gate, two-qubit gate, measurement, and shuttling. Since all of these operations are conducted by control gates, they cannot be executed simultaneously on the same electron. This constraint is shown in the following formula \ref{fsame}, where $o_{b_j}$ denotes an operation on electron $b_j$, such as $(g1, b_j)$. The same applies to $o_{b_k}$.
\begin{align}
\forall o_{b_j}, o_{b_k}, \forall b_j, b_k \in B \cdot o_{b_j} \in l_{i+1} \land o_{b_k} \in l_{i+1} \nonumber \\
\Rightarrow b_j = b_k \label{fsame}
\end{align}

\subsection{Requirements of Transition Path}\label{requirement}
\begin{figure}[htb]
\centering
\scalebox{0.55}{\includegraphics{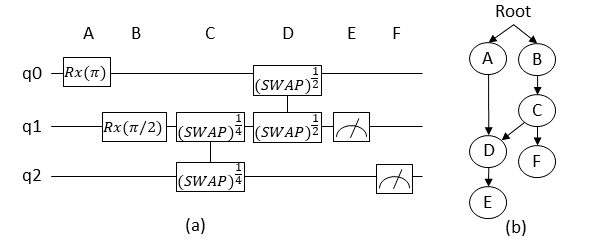}}
\caption{Example of directed acyclic graph}
\label{dag}
\end{figure}

The arbitrary transition path of $M$ is denoted by $p = [ t_{1}, ..., t_{i}, ..., t_{n} ]$, where $t_i = (s_{i-1}, l_{i}, s_{i})$. The sequence of labels $l_1, ... , l_i, ... , l_n$ represents the operation procedure of the array. If $p$ contains the operations of quantum gates and measurements in the same order as a quantum circuit, $p$ is able to execute the quantum circuit. A quantum circuit can be represented by a directed acyclic graph (DAG) whose nodes correspond to operations and whose edges indicate the order of the operations \cite{gel2012}\cite{sabre}. Figure \ref{dag}(b) represents the DAG of the quantum circuit shown in (a). For example, the $(SWAP)^{1/2}$ gate of D is executed after the $Rx(\pi)$ gate of A and $(SWAP)^{1/4}$ gate of C.
A DAG consists of $Ope$ and $Ord$. $Ope$ is the set of nodes and $Ord$ is the set of edges. A node of the DAG represents an operation. An edge from node $o_1 \in Ope$ to another node $o_2 \in Ope$ is represented as $(o_1, o_2)$, and $(o_1, o_2) \in Ord$ holds. For the target quantum circuit $QC = (Ope, Ord)$, the path $p=[ ..., (s_{i-1}, l_{i}, s_{i}), ...]$ is required to satisfy the following formulae \ref{freq1} and \ref{freq2}. 
\begin{align}
\forall o_1, o_2 \in Ope, \forall l_a, l_b \in L \cdot (o_1, o_2) \in Ord \nonumber \\
\land o_1 \in l_a \land o_2 \in l_b \Rightarrow  a < b \label{freq1}
\end{align}
\begin{align}
\forall l \in L \cdot o \in l \Rightarrow o \in Ope \label{freq2}
\end{align}
Formula \ref{freq1} expresses that the order of operations in $p$ is consistent with the order defined by the DAG. Formula \ref{freq2} means that operations that do not consist of the DAG are not included in $p$. When these formulae hold, the transition path $p$ is an operation procedure for executing the quantum circuit $QC$.

\section{Problem Statement}\label{problem}
\begin{figure*}[htb]
\centering
\scalebox{0.45}{\includegraphics{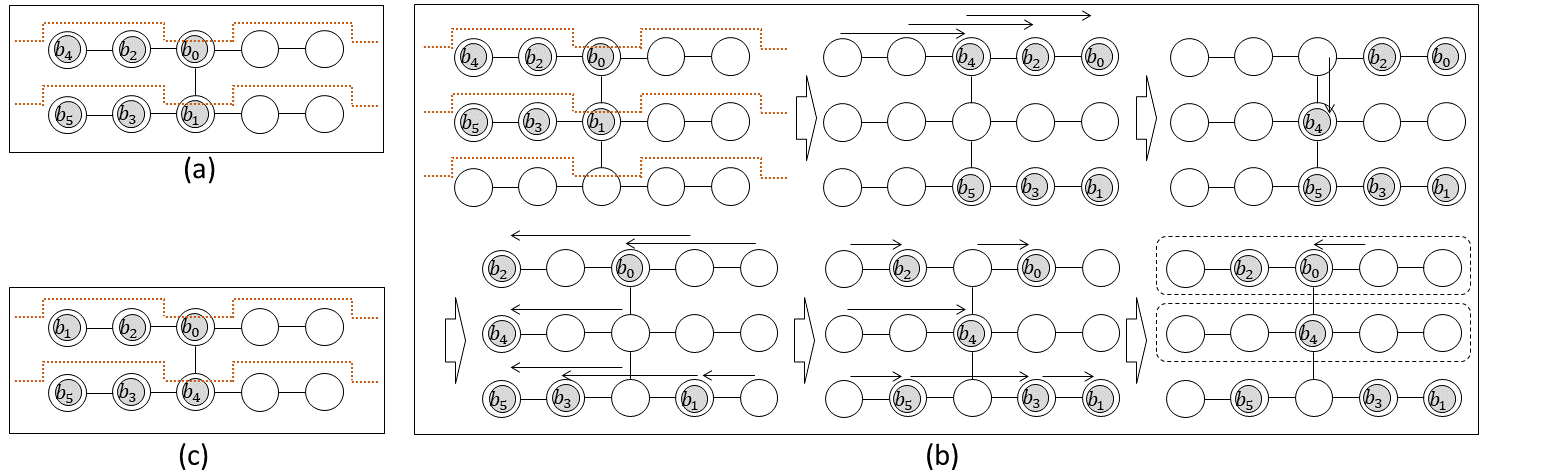}}
\caption{Example of situations with no solution}
\label{overfilling}
\end{figure*}

\begin{figure*}[htb]
\centering
\scalebox{0.45}{\includegraphics{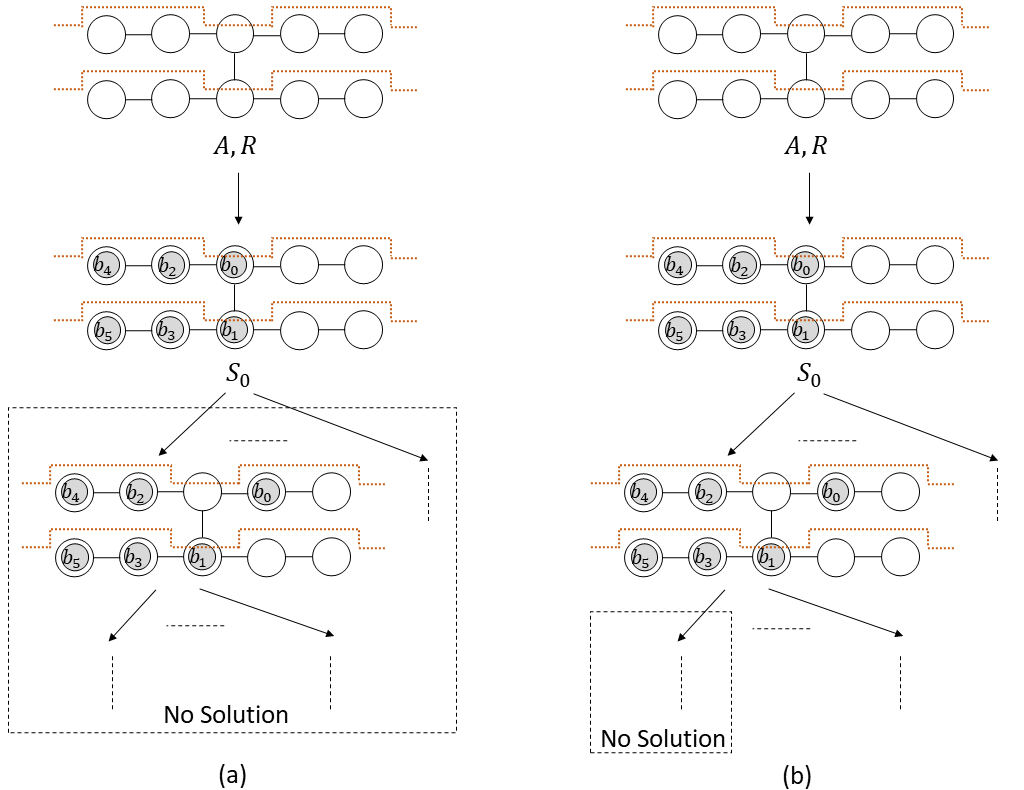}}
\caption{Problems in the search for operation procedures}
\label{problems}
\end{figure*}

When we can extract a transition path $p$ from $M$ that satisfies all the constraints and requirements discussed in Section \ref{constraint}, $p$ corresponds to a procedure to execute a quantum circuit $QC$ with SQDA. Whether $p$ satisfying the constraints and requirements can be extracted depends not only on the search method of $M$ but also on the array topology $A$, the structure of the row-shared control gates represented by $R$, and initial electron position $s_0$. Depending on $A$, $R$, and $s_0$, there may be no $p$ that satisfies the constraints and requirements. An example is shown in Fig. \ref{overfilling}.
In (a), it is not possible to execute a two-qubit gate on $b_0$ and $b_4$ because the array is overfilled with electrons relative to the number of quantum dots. If we add a third row in the array, as shown in (b), we can execute that quantum gate. Otherwise, if the target quantum circuit includes only a two-qubit gate on $b_0$ and $b_4$ but not a two-qubit gate on $b_0$ and $b_1$, then the electrons should initially be placed as in (c). This also allows the two-qubit gate on $b_0$ and $b_4$ to be executed without adding the third row. As this example shows, certain quantum circuits may not be executable depending on $A$, $R$, and $s_0$. That is, as shown in Fig. \ref{problems}(a), there may be no $p$ that satisfies the constraints and requirements in the search space of $M$. 

\begin{figure*}[htb]
\centering
\scalebox{0.45}{\includegraphics{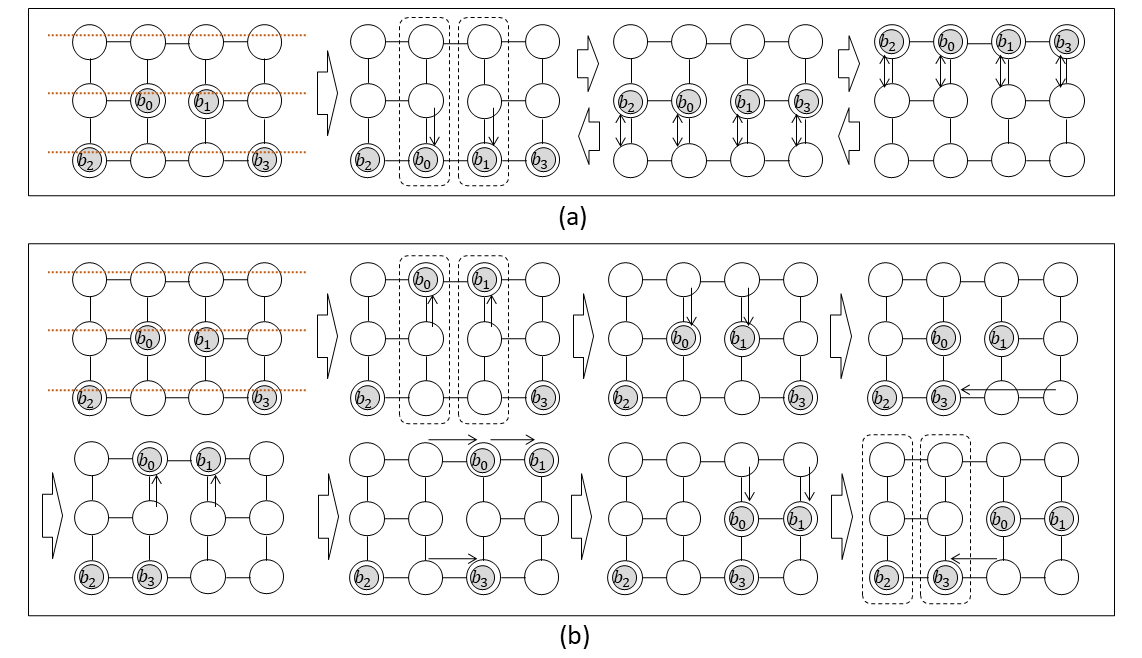}}
\caption{Example of exploring a subspace with no solution}
\label{livelock}
\end{figure*}

As the scale of quantum circuits increases, it becomes difficult to execute breadth-first search due to the limitation of memory, so $M$ is explored using a depth-first search. However, due to the complicated constraints and requirements, it is possible to search a space with no solution, as shown in Fig. \ref{problems}(b). An example is presented in Fig. \ref{livelock}.
In this example, a two-qubit gate is executed on $b_0$ and $b_1$, followed by a two-qubit gate on $b_2$ and $b_3$. The quantum dots in which $b_0$ and $b_1$ are trapped are not connected by channels. Therefore, in (a), $b_0$ and $b_1$ are shuttled to the bottom row of the array to execute a two-qubit gate. However, this results in $b_0$, $b_1$, $b_2$, and $b_3$ being in the same row. This means that only shuttling operations that move these four electrons upward or downward at the same time can be executed. For example, $b_2$ and $b_3$ cannot be adjacent to each other to conduct a two-qubit gate. On the other hand, in (b), by moving $b_0$ and $b_1$ upward to the upper quantum dot, the four electrons are avoided to line up in the same row. As a result, a two-qubit gate for $b_2$ and $b_3$ can be executed with the procedure shown in (b). This example suggests that certain operations may lead to a situation in which subsequent quantum gate operations cannot be executed. In such a case, it takes time to extract $p$ because it is necessary to exhaustively search a subspace with no solution and then move on to search another space by backtracking. Depending on the size of the subspace to be searched, the search may not be completed in a practical amount of time.

Other quantum computers (e.g., superconducting quantum computers) have a similar problem, called the qubit mapping problem, in which the constraint that the qubits to be executed in a two-qubit gate must be adjacent is called the nearest neighbor constraint (NN-constraint) \cite{Bhattacharjee2018ANA}\cite{determiningminimal}. In the qubit mapping problem, the mapping between logical and physical qubits is changed by inserting a SWAP gate so that the NN-constraint is satisfied. The SWAP gate moves the contents of the target qubits so that they are adjacent for the two-qubit gate operation. The problem of finding the optimal SWAP gate insertion to satisfy the NN-constraint is NP-hard \cite{Botea2018OnTC}\cite{Maslov2008}\cite{Siraichi2018}\cite{Tan2021}, and various methods for solving it have been proposed \cite{Alireza2014}\cite{Robert2016}\cite{bhattacharjee2017depthoptimal}\cite{determiningminimal}\cite{PAQCS}\cite{PAQCS2}\cite{tiket}\cite{Bhattacharjee2018ANA}\cite{sabre}\cite{efficientmapping}.
In contrast, in our SQDA, the complexity of the constraints makes it difficult to extract an operation procedure to execute a given arbitrary quantum circuit, which is the problem we address in this paper.


\section{Approach}\label{secapp}
The problem shown in Section \ref{problem} will not occur if we do not get into a situation where a particular quantum gate cannot be executed during the search for a transition path. Therefore, we construct $M$ so that the search can reach a state where any quantum gate operation can be executed from any intermediate state $s_{i} (i \geq 0)$ in $p$. In more detail, we restrict the transitions of $M$ to satisfy the following conditions C1 to C6.
\begin{align}
&[C1] \forall b_j \in B, \exists s \in OS \cdot (s, l, s) \in T \land l = \{ (g1, b_{j}) \} \nonumber \\
&[C2] \forall b_{j1}, b_{j2} \in B, \exists s \in OS \cdot (s, l, s) \in T \nonumber \\
& \quad \quad \land l = \{ (g2, b_{j1}, b_{j2}) \} \nonumber \\
&[C3] \forall b_j \in B, \exists s \in OS \cdot (s, l, s) \in T \land l = \{ (m, b_{j}) \} \nonumber \\
&[C4] \forall s_{rs} \in RS, s_{os} \in OS \cdot Reachable(s_{rs}, s_{os}) \nonumber \\
&[C5] s_0 \in RS \nonumber \\
&[C6] \forall s_{rs1} \in RS, \forall s \in S \exists s_{rs2} \in RS \cdot Reachable(s_{rs1}, s) \nonumber \\
& \quad \quad \Rightarrow Reachable(s, s_{rs2}) \nonumber
\end{align}

C1, C2, and C3 define the operation states ($OS$), which is the set of states of $M$. The $OS$ includes the states in which a single qubit gate for any electron, a two-qubit gate for any pair of electrons, and a measurement for any electron can be {\it independently executable}. If a particular operation is always conducted together with another operation, the executable quantum circuits are limited. In other words, {\it independently executable} means that no other operations are conducted simultaneously with that operation. C4 defines the set of states called ready states ($RS$), in which the states are reachable to all the states contained in $OS$. The predicate $Reachable(s_a, s_b)$ will be True if there exists an arbitrary transition path from state $s_a$ to $s_b$, or $s_a=s_b$. C5 means that the initial state $s_0$ is contained in $RS$. C6 gives the constraint that every state reachable from a state in $RS$ is reachable to at least one state contained in $RS$. In other words, if a search is started from a state in $RS$, it will never fall into a state that cannot be returned to the state in $RS$.

By restricting $M$ to satisfy C1 to C5, the search can reach the states where quantum gate operations and measurements can be independently executed on arbitrary electrons from a state in $RS$. Moreover, by satisfying C6, the search can return to a state in $RS$ from any state reached. Thus, if any state is reached during the search, it is possible to return from that state to a state in $RS$. From that state in $RS$, a transition path to conduct an arbitrary operation sequence of quantum gates and measurements can be extracted.
Among the components of $M$, the set of states $S$ and the set of transitions $T$ can be changed by the quantum compiler. On the other hand, the topology $A$ of SQDA and the configuration of the row-shared control gate $R$ are determined when the SQDA is manufactured and cannot be changed by the compiler. Therefore, it is necessary to design $A$ and $R$ before manufacturing the SQDA so that $M$ satisfies conditions C1 to C6, considering the behavior of the quantum compiler.

\begin{figure}[htb]
\centering
\scalebox{0.4}{\includegraphics{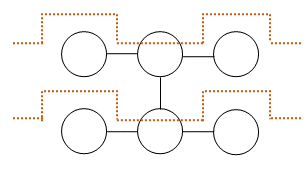}}
\caption{Small example of quantum dot array structure}
\label{smallexample1}
\end{figure}

\begin{figure*}[htb]
\centering
\scalebox{0.45}{\includegraphics{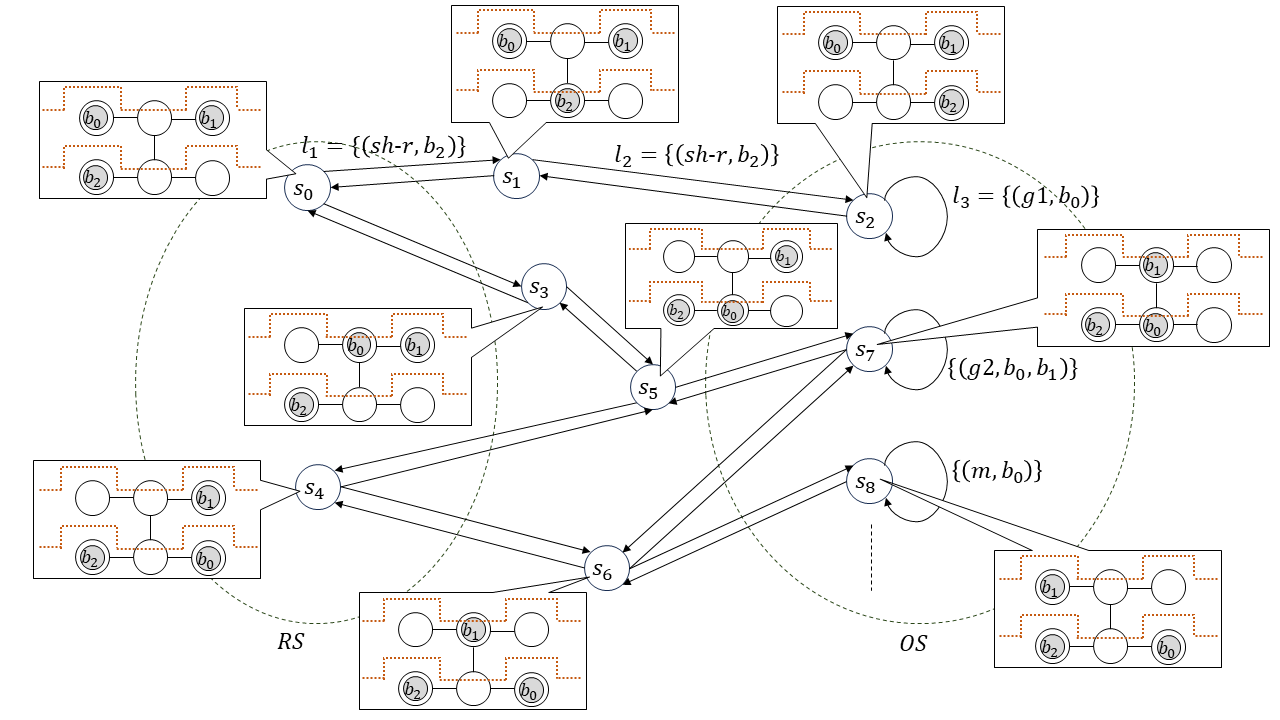}}
\caption{State transition system $M$ of the small example}
\label{smallexample2}
\end{figure*}

For example, suppose that a small array is designed as $A$ and $R$ shown in Fig. \ref{smallexample1}. If the number of qubits used in the quantum circuit to be executed is 3, i.e., $B = \{ b_0, b_1, b_2 \}$, the state transition system $M$ satisfying conditions C1 to C6 can be constructed as shown in Fig. \ref{smallexample2}.
The electrons are initially placed as shown in $s_0$. The transition path for conducting a single-qubit gate for $b_0$ is $[(s_0, l_1, s_1), (s_1, l_2, s_2), (s_2, l_3, s_2)]$. The transition from $s_0$ to $s_1$ is caused by the shuttling of electron $b_2$ to the quantum dot on the right, so $l_1 = \{ (sh \text{-} r, b_2) \}$. In state $S_2$, the loop transition representing the execution of a single-qubit gate occurs, so $l_3 = \{ (g1, b_0) \}$.

Figure \ref{smallexample2} shows only three states in $OS$ as examples---single-qubit gate for $b_0$ ($s_2$), two-qubit gates for $b_0$ and $b_1$ ($s_7$), and measurement for $b_0$ ($s_8$)---but we assume that operations on other electrons are also included in $OS$. Thus, C1, C2, and C3 are satisfied. C4 holds because the states $s_0$ and $s_4$ contained in $RS$ are reachable to $s_2$, $s_7$, and $s_8$ contained in $OS$. C5 is also satisfied since $s_0$ is contained in $RS$. For $s_1$, $s_2$, $s_3$, $s_5$, $s_6$, $s_7$, and $s_8$ that are reachable from $s_0$ and $s_4$, they are reachable to either $s_0$ or $s_4$ contained in $RS$. Therefore, C6 is held.

\section{State Transition System for 16 $\times$ 8 SQDA}\label{instance}

We present a concrete state transition system $M$ for our 16 $\times$ 8 SQDA. The transitions of $M$ should satisfy conditions C1 to C6 and the constraints represented by formulae \ref{f1qubit} to \ref{fsame}
\subsection{Array Structure}
We define the structure of SQDA consisting of 16 columns x 8 rows of quantum dots as shown in Fig. \ref{electronposition}. The structure is defined by $A=(V, E)$ representing the array topology and $R$ representing the row-shared control gate. They satisfy the following formulae \ref{st-1} to \ref{st-5}.
\begin{align}
\forall v_{r,c} \in V, \exists n \in \mathbb{N} \cdot c = 2n \Rightarrow R(v_{r,c}) \label{st-1}
\end{align}
\begin{align}
\forall v_{r,c} \in V, \exists n \in \mathbb{N} \cdot c = 2n - 1 \Rightarrow \lnot R(v_{r,c}) \label{st-2}
\end{align}
\begin{align}
\forall v_{r1,c1}, v_{r2,c2} \in V, \exists n \in \mathbb{N} \cdot c1 = 2n \land c1 = c2 \nonumber \\
\land |r1 - r2| = 1 
\Rightarrow (v_{r1,c1}, v_{r2,c2}) \in E \label{st-3}
\end{align}
\begin{align}
\forall v_{r1,c1}, v_{r2,c2} \in V, \exists n \in \mathbb{N} \cdot c1 = 2n - 1 \land c1 = c2  \nonumber \\
\land |r1 - r2| = 1 \Rightarrow (v_{r1,c1}, v_{r2,c2}) \notin E \label{st-4}
\end{align}
\begin{align}
\forall v_{r1,c1}, v_{r2,c2} \in V \cdot |c1 - c2| = 1 \land r1 = r2 \nonumber \\
\Rightarrow (v_{r1,c1}, v_{r2,c2}) \in E \label{st-5}
\end{align}
$\mathbb{N}$ is the set of natural numbers. Formula \ref{st-1} represents that even-numbered columns of quantum dots are connected to row-shared control gates. Similarly, formula \ref{st-2} represents that the quantum dots in odd-numbered columns are not connected to the row-shared control gate. Thus, by connecting the row-shared control gate every other column, block control is possible in all columns based on formula \ref{fbccondition}.
Formula \ref{st-3} means that the vertically adjacent quantum dots in even-numbered columns are connected by channels. Formula \ref{st-4} indicates that vertically adjacent quantum dots in odd-numbered columns are not connected by channels. Formula \ref{st-5} expresses that all horizontally adjacent quantum dots are connected by channels. Thus, by not installing vertical channels in odd-numbered columns, even if electrons in even-numbered columns are moved vertically, the electrons placed in odd-numbered columns are not moved together with them from formulae \ref{fblockctrl3} and \ref{fblockctrl4}.

\subsection{Ready States}
We define ready states $RS$ based on the above structure. A state in which all electrons are placed in a quantum dot that is not connected to a row-shared control gate is included in $RS$. In those states in $RS$, quantum dots in which electrons can be placed are called {\it seat} dots.
No electrons are placed in the quantum dots of a particular row. This specific row is called the {\it bus} row and its row number is denoted as $r_{bus}$. The bus row is utilized as a pathway for moving electrons to another column to execute a two-qubit gate. By using block control, electrons can be moved through the bus row while other electrons remain in place. The columns connected to the row-shared control gate are also used as pathways for electrons to exit the bus row. These columns are called {\it aisle} columns. The state $s_i$ satisfying the following formula \ref{seatdot} is contained in $RS$.
\begin{align}
\forall v_{r,c} \in V \cdot v_{r,c} \in P_i 
\Rightarrow \lnot R(v_{r,c}) \land r \neq r_{bus} \label{seatdot}
\end{align}

\begin{figure*}[htb]
\centering
\scalebox{0.45}{\includegraphics{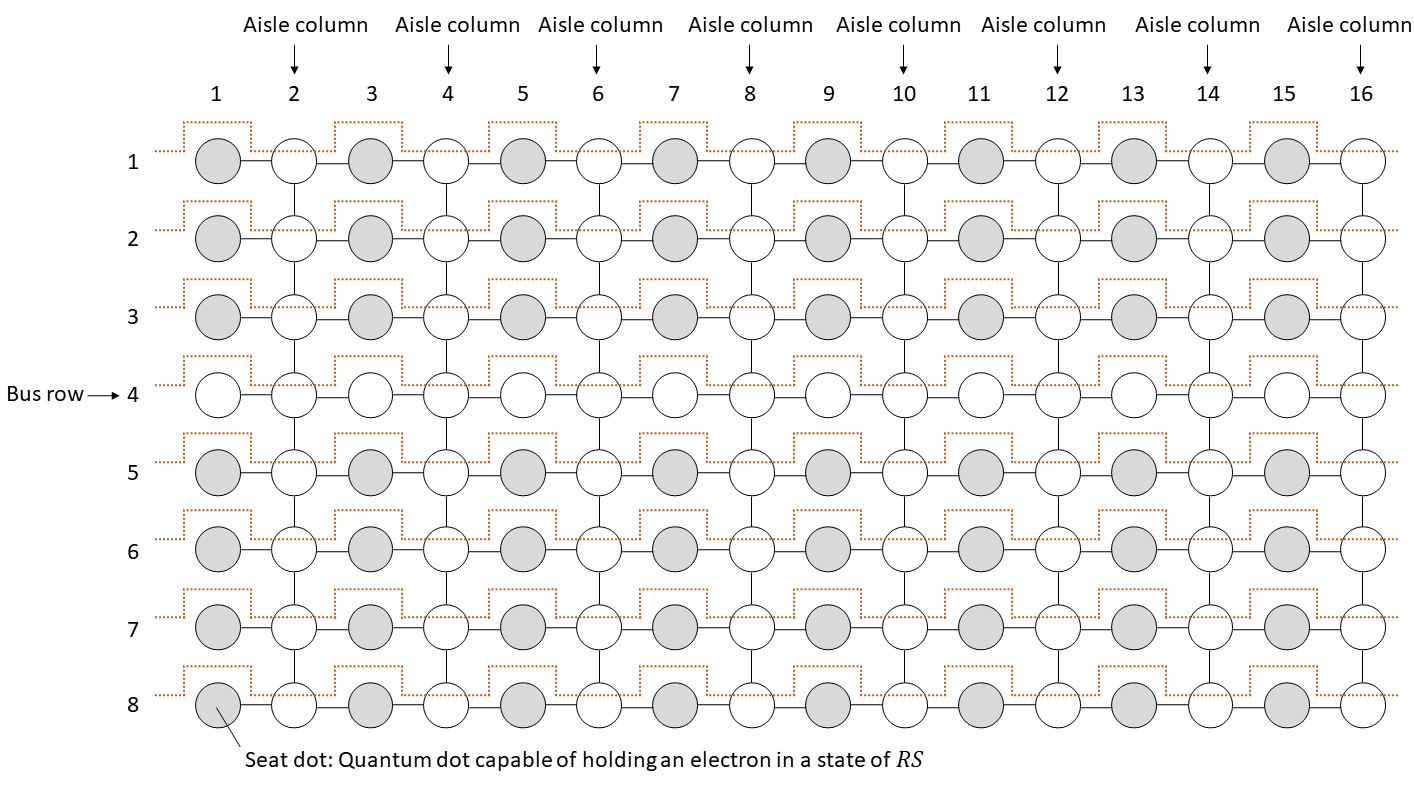}}
\caption{Quantum dots where electrons can be placed in the states of $RS$}
\label{electronposition}
\end{figure*}

\begin{figure}[htb]
\centering
\scalebox{0.18}{\includegraphics{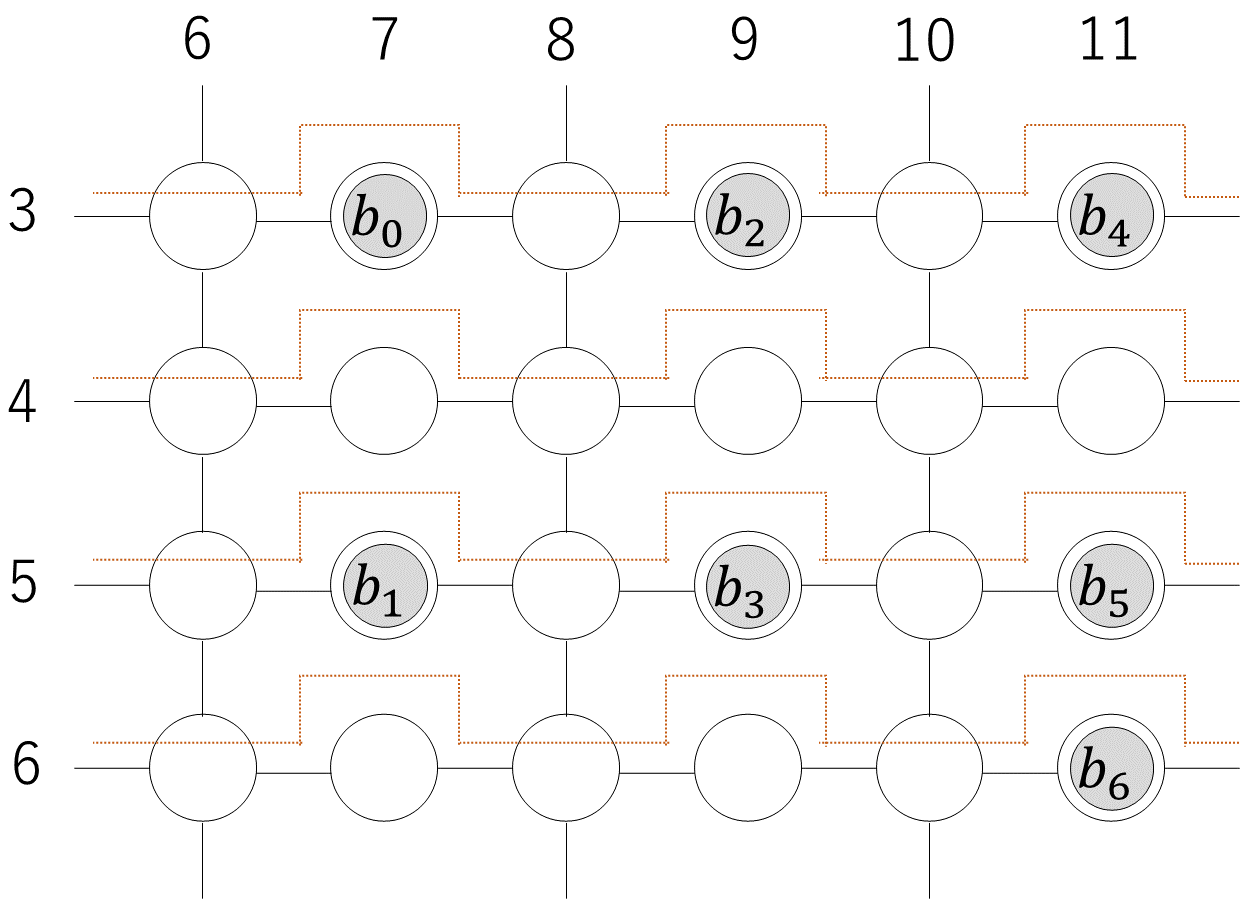}}
\caption{Example of a state included in $RS$}
\label{rsexample}
\end{figure}

In Fig. \ref{electronposition}, the quantum dots filled in gray are the seat dots when $r_{bus}=4$. If electrons are placed only in these quantum dots in $s_i$, $s_i$ is included in $RS$. Thus, for our 16 $\times$ 8 SQDA, 56 electrons can be used at most.
Figure \ref{rsexample} shows an example of a state included in $RS$ when the number of electrons is 7. All electrons exist in seat dots.

\subsection{Transitions of Single-qubit Gates}\label{sec-trans1qubit}

\begin{figure*}[htb]
\centering
\scalebox{0.29}{\includegraphics{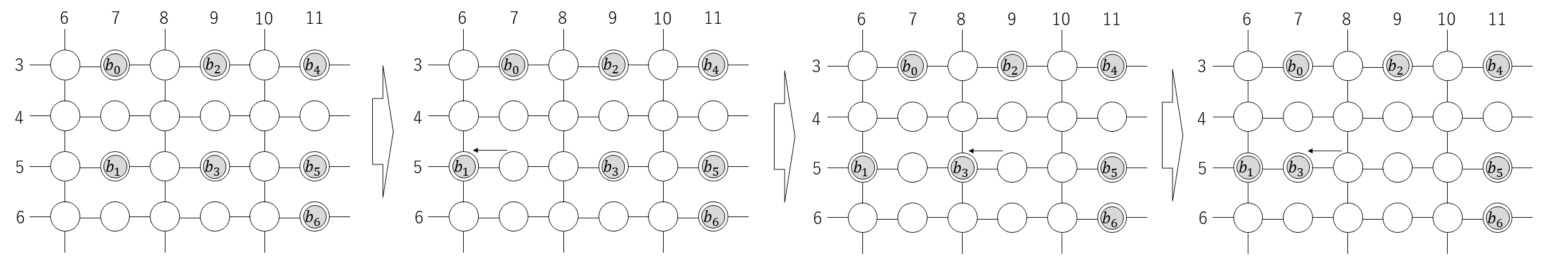}}
\caption{Example of transitions for the execution of a single-qubit gate}
\label{trans1qubitex}
\end{figure*}

We define the transitions from a state in $RS$ to a state in which a single-qubit gate can be executed independently. Let $b_j$ be the target qubit of the single-qubit gate. In order to execute a single-qubit gate for $b_j$ independently, from formula \ref{f1qubit}, it is necessary to evacuate other electrons in the column $c_j$ where $b_j$ is located. Furthermore, according to formula \ref{f1gatecross}, it is necessary to evacuate electrons outside of the adjacent columns to avoid crosstalk.
As an example, the transition from the state of $RS$ shown in Fig. \ref{rsexample} to the state where a single-qubit gate on $b_2$ is executed is shown in Fig. \ref{trans1qubitex}.
In order to execute a single qubit gate for $b_2$ in column 9, $b_3$ in the same column needs to be moved out. Therefore, $b_1$ in column 7 is first moved to column 6 to create the space for $b_3$, Next, $b_3$ is moved to column 7. As a result, there are no other electrons in the same and adjacent columns of $b_2$. This avoids crosstalk and allows a single-qubit gate to be executed independently. In this example, $b_3$ is evacuated to the left side, but it can also be evacuated to the right column 11.

\begin{figure}[htb]
\centering
\scalebox{0.45}{\includegraphics{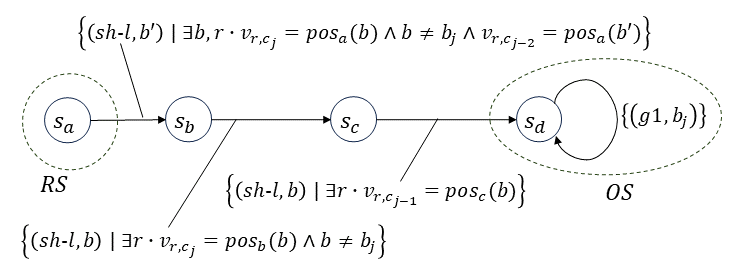}}
\caption{Transitions to execute a single-qubit gate}
\label{trans1qubit}
\end{figure}

The transitions for executing a single-qubit gate operation on an arbitrary electron $b_j$ independently are shown in Fig. \ref{trans1qubit}. The column where $b_j$ is located in state $s_a$ is represented by $c_j$. That is, $v_{r_{j},c_{j}} = pos_a(b_j)$. This figure shows only the transitions for the case where electrons in the same column as $b_j$ are evacuated to the left side. A similar transition can be performed for the case of evacuating to the right side. If $c_j \leq 3$, it cannot be evacuated to the left, so it is evacuated to the right. Similarly, if $c_j \geq 14$, it is moved to the left side.
In the transition from $s_a$ to $s_b$, electrons in column $c_{j}-2$ are shuttled to the left to create an evacuation area for electrons other than $b_j$ in column $c_j$. In rows where no electrons are located in column $c_j$, electrons in column $c_{j}-2$ need not be evacuated. Therefore, only electrons in row $r$ that satisfy $v_{r,c_j}=pos_a(b)$---that is, only electrons in the same row $r$ as $b$, which is in the same column as $b_j$---are moved. Next, in the transition from $s_b$ to $s_d$, all electrons in column $c_{j}$ except $b_j$ are shuttled left to column $c_{j}-2$. These transitions satisfy the shuttling constraints represented by formulae \ref{fshconst1} to \ref{fsame}. The transition $(s_d, l, s_d) in T$ representing a single-qubit gate operation, where $l = \{ (g1, b_{j}) \}$, is performed in $s_d$. This transition satisfies the constraint of formula \ref{f1qubit}. In addition, since there are no electrons in the adjacent columns, the constraint represented by formula \ref{f1gatecross} is also satisfied. Therefore, crosstalk can be avoided.

We next define transitions from state $s_d$ to return to a state included in $RS$. We use the reverse procedure of transitioning from $s_a$ to $s_d$.
\begin{figure}[htb]
\centering
\scalebox{0.45}{\includegraphics{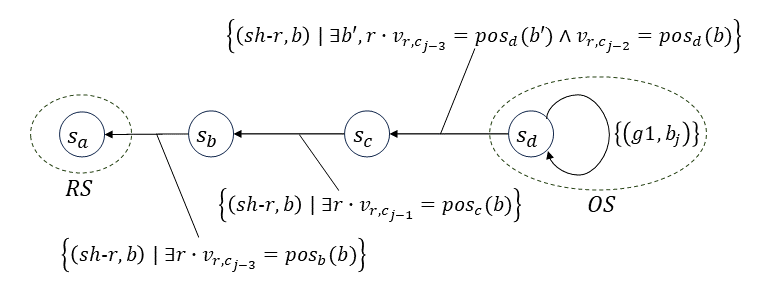}}
\caption{Transitions to return to a state in $RS$ after the single-qubit gate operation}
\label{return1qubit}
\end{figure}
Figure \ref{return1qubit} represents the transitions back to $s_a$ in $RS$ after the electron in the same row as $b_j$ is evacuated to the left as shown Fig. \ref{trans1qubit}. In the transitions from $s_d$ to $s_b$, the electrons, which are evacuated in column $c_{j}-2$, are returned to the original column $c_{j}$. Then, in the transition from $s_b$ to $s_a$, the electrons in column $c_{j}-3$ are returned to the original column $c_{j}-2$. These transitions satisfy the shuttling constraints represented by formulae \ref{fshconst1} to \ref{fsame}. The same transitions can be performed to return to the state in $RS$ when the electrons are evacuated to the right side.

\subsection{Transitions of Two-qubit Gates}\label{sec-trans2qubit}

\begin{figure*}[htb]
\centering
\scalebox{0.4}{\includegraphics{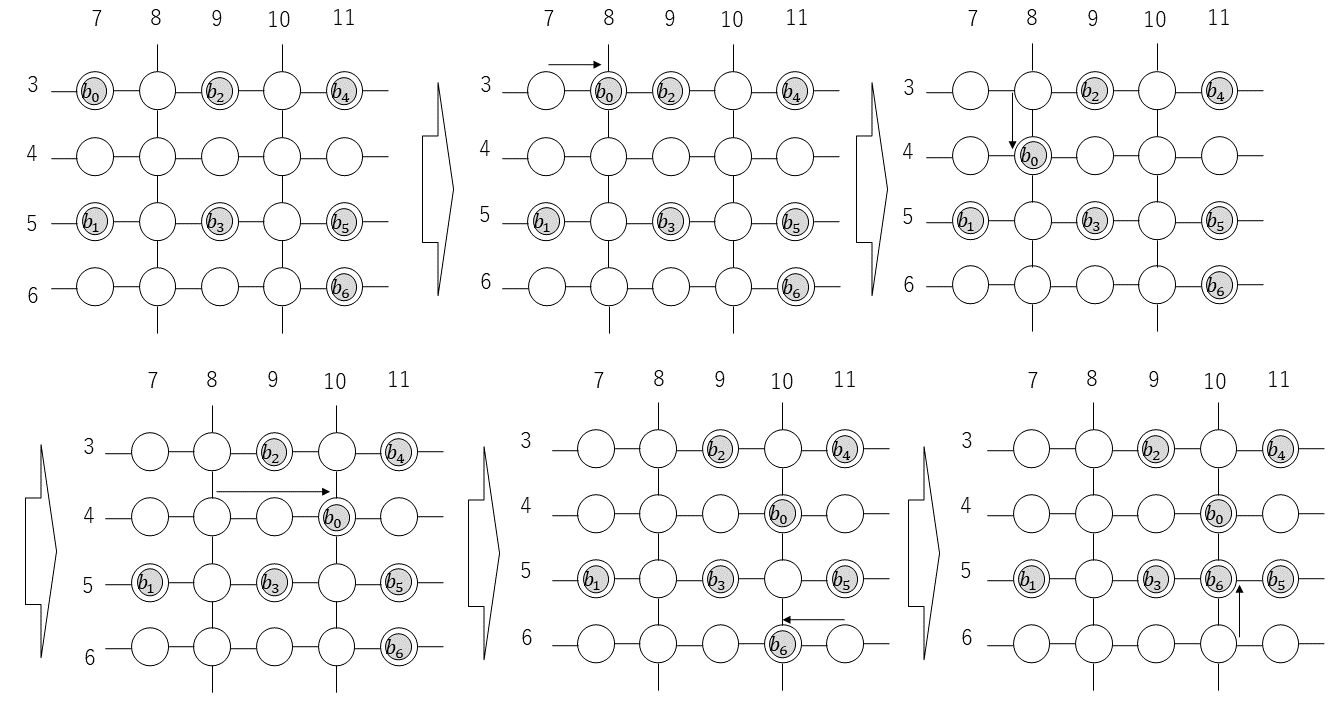}}
\caption{Example of transitions for the execution of a two-qubit gate}
\label{trans2qubitex}
\end{figure*}

\begin{figure*}[htb]
\centering
\scalebox{0.4}{\includegraphics{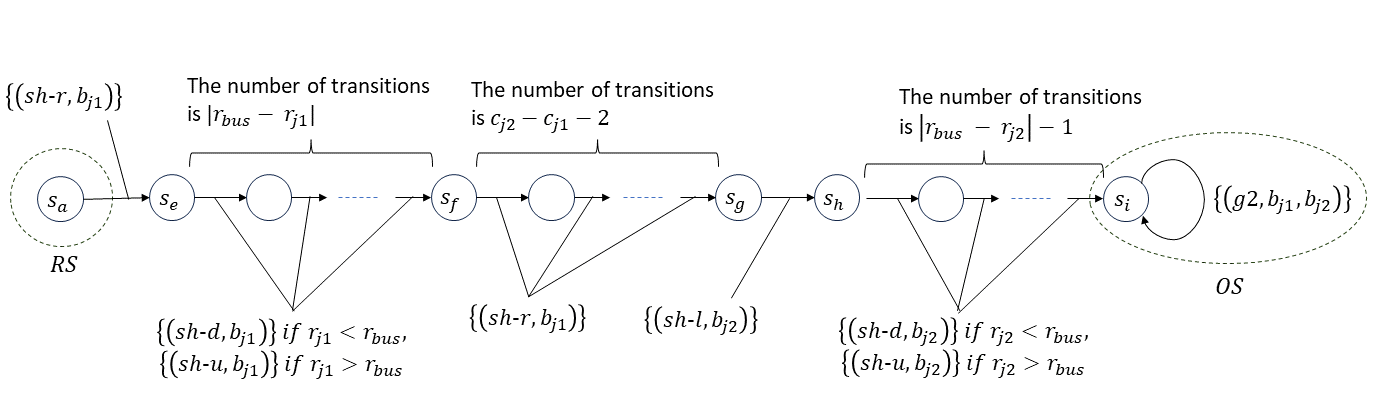}}
\caption{Transitions to execute a two-qubit gate}
\label{trans2qubit}
\end{figure*}

We define the transitions from state $s_a$ in $RS$ to a state in which the two-qubit gate is executed independently. Let $b_{j1},b_{j2}$ be the target electrons and $v_{r_{j1},c_{j1}} = pos_a(b_{j1})$ and $v_{r_{j2},c_{j2}} = pos_a(b_{j2})$. Assume $c_{j1} \leq c_{j2}$ without loss of generality. 
We execute a two-qubit gate by vertically adjoining $b_{j1}$ and $b_{j2}$ and using row-sharing control gates. In a two-qubit gate operation, the same operation may be conducted on electrons in the same row (formula \ref{f2qubitver}) or may be affected by the operation (formula \ref{fcross2qubitver}). In any states in $RS$, none of the seat dots where electrons are located are vertically connected by a channel (Fig. \ref{electronposition}). Therefore, if no electrons other than $b_{j1}$ and $b_{j2}$ are moved, a two-qubit gate can be conducted on $b_{j1}$ and $b_{j2}$ without effects. Fig. \ref{trans2qubitex} shows an example of the transition from the state shown in Fig. \ref{rsexample} to a state to execute a two-qubit gate on $b_0$ and $b_6$.
The electron $b_{0}$ moves through the adjacent aisle column to the bus row. It then passes through the bus row to the aisle column adjacent to $b_{6}$. Next, $b_{6}$ moves through the adjacent aisle column to the quantum dot vertically adjacent to $b_{0}$.
Figure \ref{trans2qubit} shows the transitions to execute a two-qubit gate on $b_{j1},b_{j2}$.
The transition from state $s_a$ to $s_e$ moves the electron $b_{j1}$ to the left aisle column, and the transitions from $s_e$ to $s_f$ transfer $b_{j1}$ through the aisle column to reach the quantum dot that intersects the bus row. The transition from $s_f$ to $s_g$ moves $b_{j1}$ through the bus row to the quantum dot where the bus row intersects the aisle column to the left of the row where $b_{j2}$ is located. However, if the aisle column to the right of $b_{j1}$ and the aisle column to the left of $b_{j2}$ are the same ($c_{j2}-c_{j1}=2$), the transition from $s_f$ to $s_g$ is not necessary. Similarly, if electrons $b_{j1},b_{j2}$ are in the same column, we use the aisle column adjacent to the right of them as the meeting point. Therefore, the transition from $s_f$ to $s_g$ is unnecessary.
The transition from $s_g$ to $s_h$ moves electron $b_{j2}$ to the left aisle column (when $b_{j1},b_{j2}$ are in the same column, they are exceptionally moved to the right aisle column, that is, the label of the transition from $s_g$ to $s_h$ is $\{ (sh \text{-} r, b_{j2}) \}$ ). Since $b_{j1}$ is waiting at the quantum dot where the aisle column and the bus row intersect, the transition from $s_h$ to $s_i$ moves $b_{j2}$ to a position adjacent to $b_{j1}$. These transitions satisfy the shuttling constraints represented by formulae \ref{fshconst1} to \ref{fsame}. The transition $(s_i, l, s_i) \in T$ representing a two-qubit gate operation, where $l = \{ (g2, b_{j1}, b_{j2} \}$, is executed in $s_i$. This transition satisfies the constraint of formula \ref{f2qubitver}. In addition, since the electrons in the same row as $b_{j1}$ and $b_{j2}$ are located in the seat dots that are not connected to the neighboring dots by channels, they are not affected by the two-qubit gate operation. This means that the constraint represented by formula \ref{fcross2qubitver} is satisfied. In Fig. \ref{trans2qubit}, $b_{j1}$ is moved to the side of $b_{j2}$, but a two-qubit gate can also be executed by moving $b_{j2}$ to the side of $b_{j1}$ by a similar procedure. Therefore, we also define the transitions for moving $b_{j2}$ to $b_{j1}$ in $M$.

\begin{figure*}[htb]
\centering
\scalebox{0.45}{\includegraphics{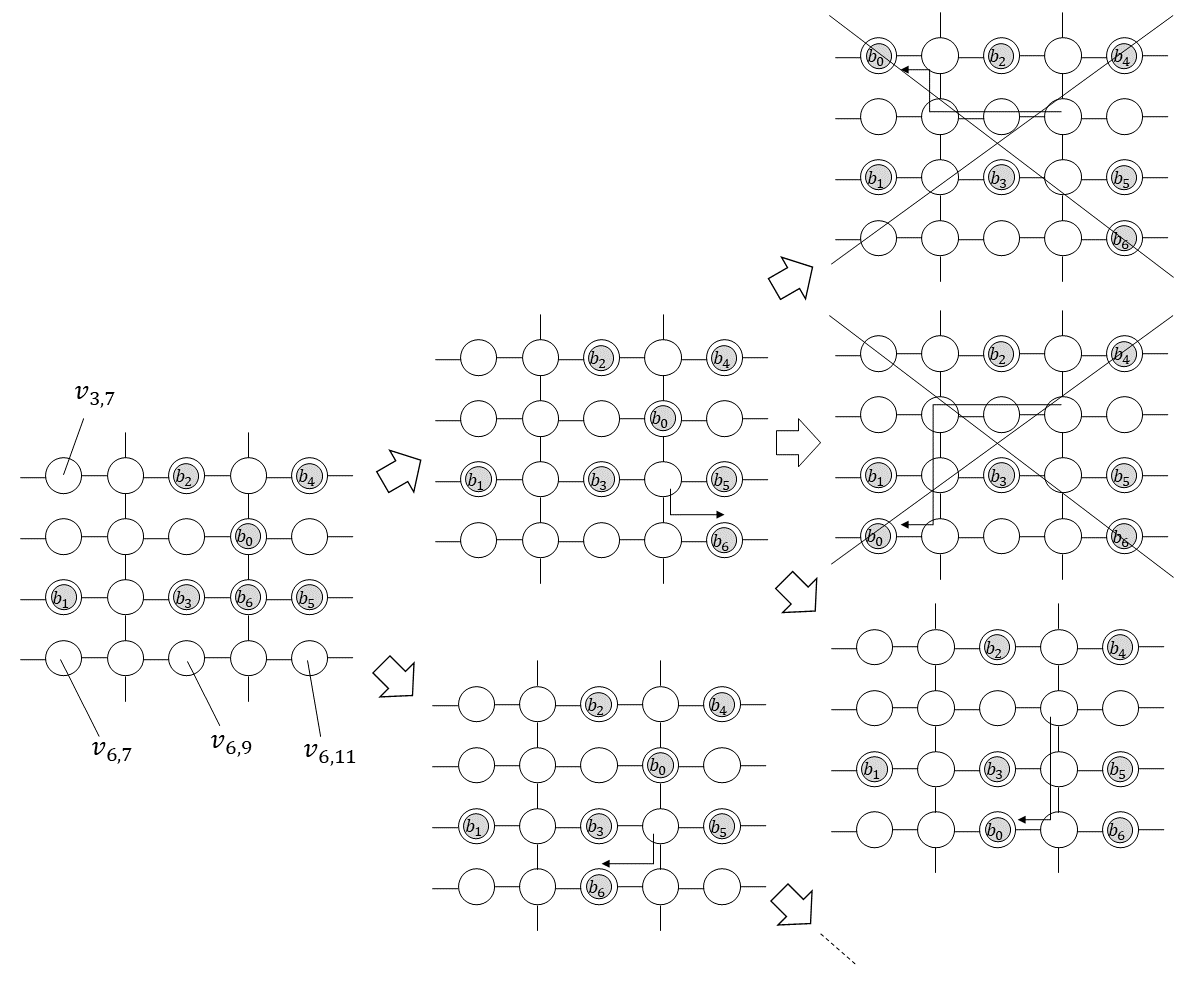}}
\caption{Example of transitions after the two-qubit gate operation}
\label{return2qubitex}
\end{figure*}

Next, we give the transitions to return to a state in $RS$ after executing the two-qubit gate. The electrons that the two-qubit gate were executed on do not have to be returned to the original seat dots. If there is another empty seat dot with a closer distance (distances in this paper are Manhattan distances), it is better to return to that seat dot to reduce the number of shuttling operations. Figure \ref{return2qubitex} shows an example.
The seat dot where $b_{6}$ was originally located is $v_{6,11}$. If there is an empty seat dot closer than $v_{6,11}$ from $b_{6}$, $b_{6}$ should be moved to that seat dot. The distance from $b_{6}$ to $v_{6,9}$ is the same as the distance to $v_{6,11}$. In this case, $b_{6}$ can move to either $v_{6,9}$ or $v_{6,11}$.
Since $b_{0}$ exists in the bus row, $b_{6}$ cannot move through the bus row to the seat dots on the other side of the bus row. Therefore, to move to the seat dot on the other side of the bus row, it has to take a different route. For example, it is possible to reach the other side of the bus row by a detour route through $v_{6,9}$. However, a detour route should always pass through a seat dot such as $v_{6,9}$. The distance to move through $v_{6,9}$ to another seat dot is obviously longer than the distance to $v_{6,9}$. Therefore, when returning electrons to the nearest seat dot, no detour route needs to be considered.
Suppose $b_{6}$ moves to $v_{6,11}$. Then, if $b_{0}$ moves to an empty seat dot, it can return to the state in $RS$. As empty seat dots that $b_{0}$ can move, $v_{3,7}$, $v_{6,7}$, and $v_{6,9}$ are candidates. Since the shortest distance to move is $v_{6,9}$, $b_{0}$ moves to $v_{6,9}$.

\begin{figure*}[htb]
\centering
\scalebox{0.42}{\includegraphics{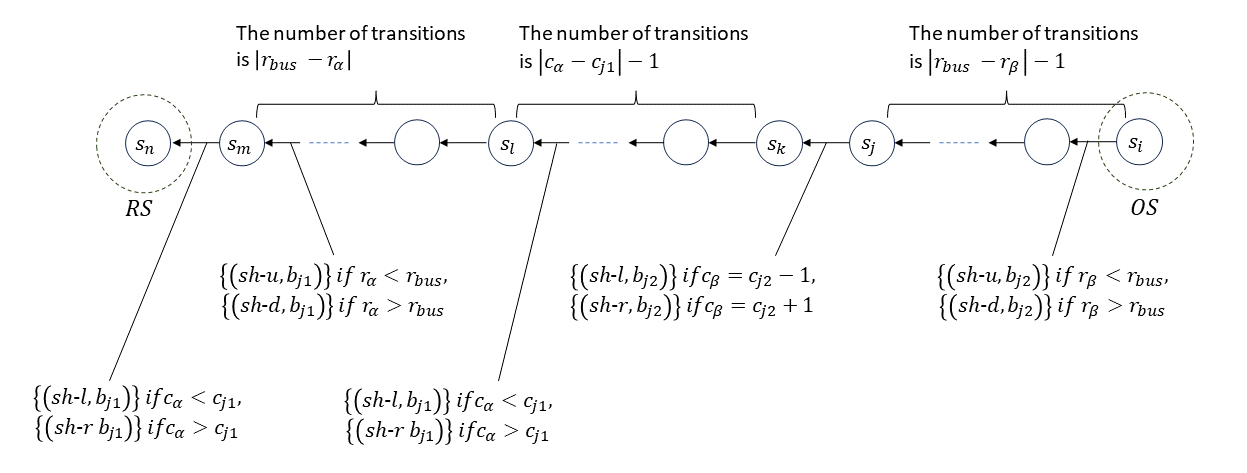}}
\caption{Transitions to return to a state in $RS$ after the two-qubit gate operation}
\label{return2qubit}
\end{figure*}

Figure \ref{return2qubit} shows the transitions to return to a state in $RS$ after executing the two-qubit gate. Let $v_{r_\beta,c_\beta}$ be the quantum dot to which $b_{j2}$ can return, which satisfies the following formula \ref{betacond}. Electron $b_{j2}$ moves to the nearest quantum dot that satisfies this condition. The positions of $b_{j1}$ and $b_{j2}$ in state $s_i$ are represented by $r_{j1},c_{j1}$ and $r_{j2},c_{j2}$, respectively. That is, $v_{r_{j1},c_{j1}} = pos_i(b_{j1})$ and $v_{r_{j2},c_{j2}} = pos_i(b_{j2})$.
\begin{align}
& (r_{j2} = r_{bus} - 1 \Rightarrow 1 \leq r_\beta < r_{bus}) \nonumber \\
& \land (r_{j2} = r_{bus} + 1 \Rightarrow r_{bus} < r_\beta \leq ROW) \nonumber \\
& \land (c_\beta = c_{j2} - 1 \lor c_\beta = c_{j2} + 1) \nonumber \\
& \land v_{r_\beta,c_\beta} \notin P_i \label{betacond}
\end{align}
In Fig. \ref{return2qubit}, the transitions from $s_i$ to $s_k$ represent a move of $b_{j2}$ to $v_{r_\beta,c_\beta}$. In the transitions from $s_i$ to $s_j$, $b_{j2}$ moves through the aisle column to the quantum dot adjacent to $v_{r_\beta,c_\beta}$. Then, $s_k$ moves $b_{j2}$ to $v_{r_\beta,c_\beta}$ by the transition to $s_k$.
Similarly, let $v_{r_\alpha,c_\alpha}$ be the quantum dot to which $b_{j1}$ can return; then, $v_{r_\alpha,c_\alpha}$ satisfies the following formula \ref{alphacond}.
The $b_{j1}$ moves to the nearest quantum dot that satisfies this condition.
\begin{align}
 1 \leq r_\alpha \leq ROW \land 1 \leq c_\alpha \leq COL \land \lnot R(v_{r_\alpha,c_\alpha}) \nonumber \\
\land r_\alpha \neq r_{bus} 
\land v_{r_\alpha,c_\alpha} \notin P_i \label{alphacond}
\end{align}
In the transition from $s_k$ to $s_l$, $b_{j1}$ moves through the bus row to the aisle column adjacent to $v_{r_\alpha,c_\alpha}$ . The transition from $s_l$ to $s_m$ moves $b_{j1}$ through the aisle column to the quantum dot adjacent to $v_{r_\alpha,c_\alpha}$. Then, in the transition to $s_n$, $b_{j1}$ moves to $v_{r_\alpha,c_\alpha}$. These returning transitions satisfy the shuttling constraints represented by formulae \ref{fshconst1} to \ref{fsame}.

\subsection{Transitions of Measurements}\label{sec-transmeasure}
\begin{figure*}[htb]
\centering
\scalebox{0.35}{\includegraphics{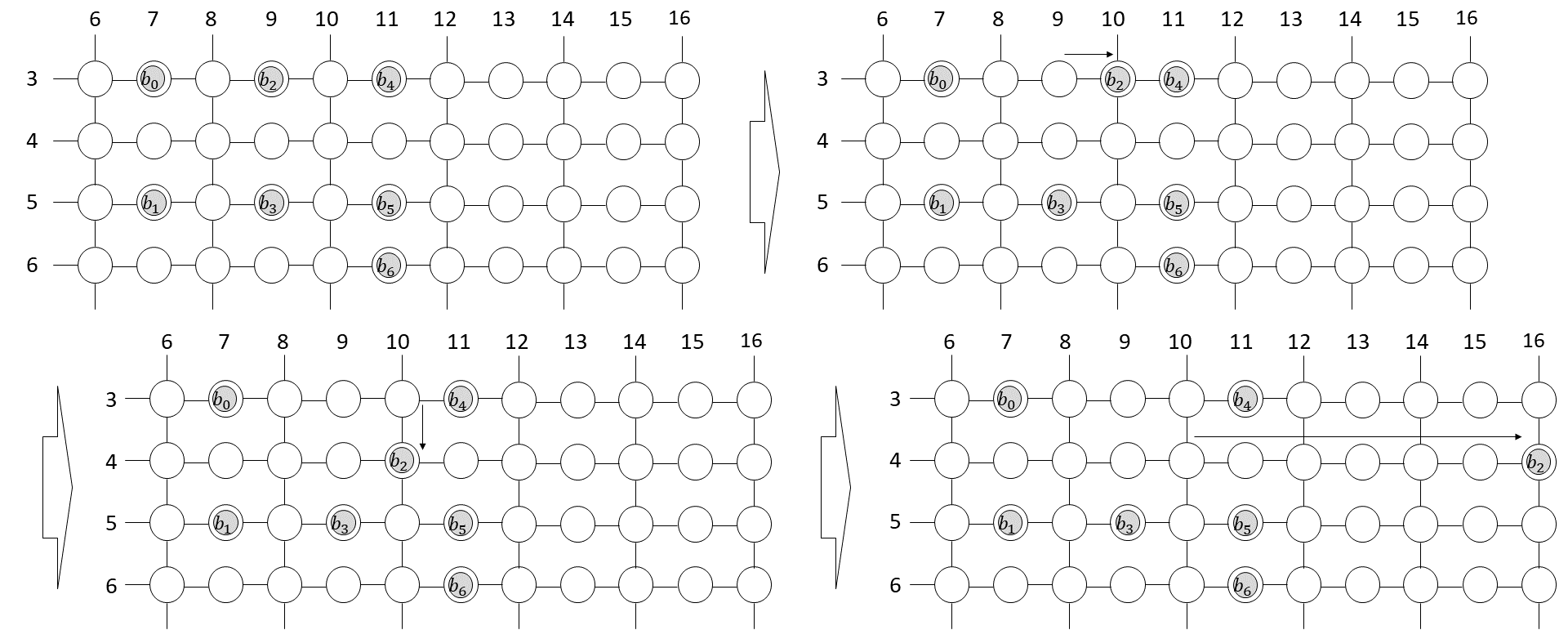}}
\caption{Example of transitions for the execution of a measurement}
\label{transmeasureex}
\end{figure*}

We define the transitions from state $s_a$ in $RS$ to a state in which the measurement is executed independently. Let $b_j$ be the qubit to be measured, where $v_{r_{j},c_{j}} = pos_a(b_j)$. From formula \ref{fmeasure2}, $b_j$ should be moved to the rightmost column $COL=16$ for the execution of the measurement. As an example, Fig. \ref{transmeasureex} shows the transitions for the measurement on $b_2$.
When executing the measurement, it is necessary to evacuate other electrons in the same column, from formula \ref{fmeasure2}. In states in $RS$, no electrons are placed in column 16 where the measurement can be conducted, so no shuttling operation is required to evacuate the electrons from column 16.

\begin{figure}[htb]
\centering
\scalebox{0.38}{\includegraphics{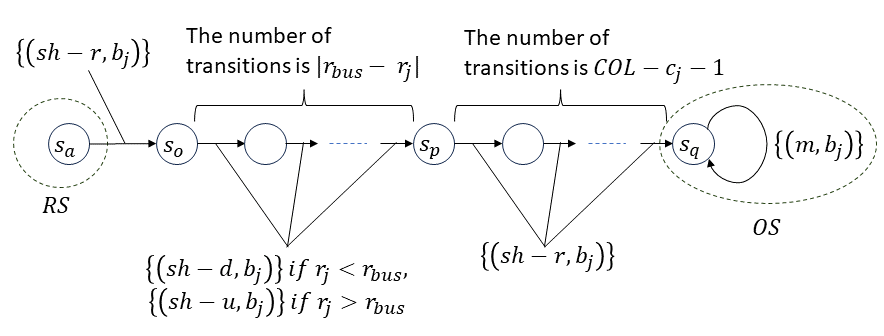}}
\caption{Transitions to execute a measurement}
\label{transmeasure}
\end{figure}

The transitions from $s_a$ of $RS$ for executing the measurement for $b_j$ are shown in Fig. \ref{transmeasure}. As exception cases, when the target electron is in column 15, it can be moved to column 16 simply by shuttling to the right. The transitions of such cases are shown in Appendix \ref{transmeasure-exception}.
The transitions from state $s_a$ to $s_p$ are similar to those for a two-qubit gate. The electron $b_{j}$ passes through the right-neighboring aisle column and reaches the quantum dot where the aisle column and the bus row intersect. Then, in the transition from $s_p$ to $s_q$, $b_{j}$ goes through the bus row and reaches the quantum dot in the rightmost column 16. These transitions obviously satisfy the shuttling constraints represented by formulae \ref{fshconst1} to \ref{fsame}. The transition $(s_q, l, s_q) \in T$ representing a measurement operation, where $l = \{ (m, b_{j}) \}$, is executed in $s_q$. This transition satisfies the constraints expressed by formulae \ref{fmeasure1} and \ref{fmeasure2}.

\begin{figure}[htb]
\centering
\scalebox{0.4}{\includegraphics{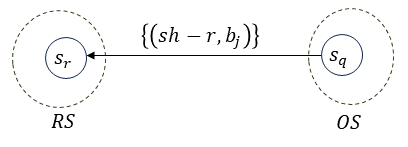}}
\caption{Transition to return to a state in $RS$ after the measurement operation}
\label{returnmeasure}
\end{figure}

We define the transition to return to a state in $RS$ after the measurement is executed. The electron $b_{j}$ measured in the rightmost column of the array is ejected to the reservoir region connected to the right of the array by shuttling to the right \cite{Lee_2022}. That is, after the measurement, $b_{j}$ disappears from the array. Since no electrons other than $b_{j}$ are moved, the state returns to a state in $RS$ when $b_{j}$ disappears. The transition back to the state in $RS$ after the measurement is shown in Fig. \ref{returnmeasure}.
In $s_q$, the electron $b_{j}$ is located in the rightmost column 16 and the other electrons are located in the seat dots. Therefore, when $b_{j}$ disappears from the array by shuttling it to the right, the state transits to state $s_r$ that is included in $RS$. The returning transition also satisfies the shuttling constraints represented by formulae \ref{fshconst1} to \ref{fsame}.

\subsection{Overview of All Transitions}\label{overviewtransition}
\begin{figure*}[htb]
\centering
\scalebox{0.4}{\includegraphics{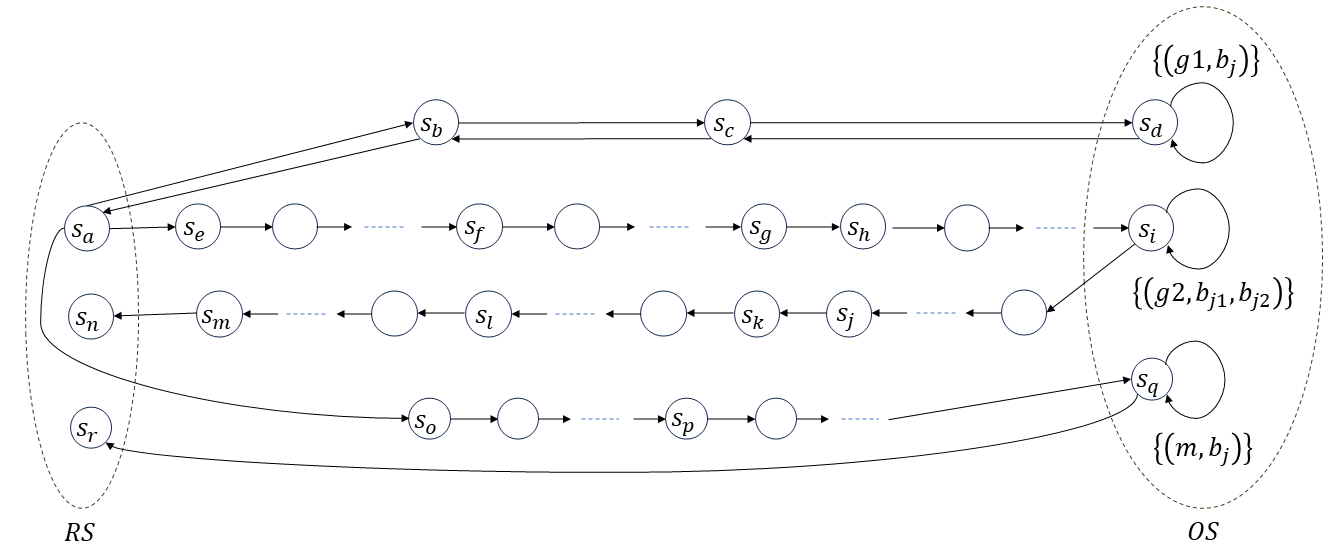}}
\caption{Overview of all transitions}
\label{overalltransitions}
\end{figure*}

The transitions defined in Sections \ref{sec-trans1qubit} to \ref{sec-transmeasure} are summarized in Fig. \ref{overalltransitions}. In $s_d$, $s_i$, and $s_q$, a single-qubit gate, two-qubit gate, and measurement, respectively, can be independently executed for any electron. Therefore, $OS$ composed of those states satisfies C1, C2, and C3. Also, $M$ satisfies C4 because it can transit from any state $s_a$ in $RS$ to those states. Although not shown in Fig. \ref{overalltransitions}, transitions to $s_d$, $s_i$, and $s_q$ are also possible from $s_n$ and $s_r$.
Since $M$ is defined such that the sequence of the transitions from $s_a$ to $s_d$ is a single path, it is guaranteed that transitions to $s_d$ are possible from all states on the path from $s_a$ to $s_d$. The same is true for transitions from $s_a$ to $s_i$, and from $s_a$ to $s_q$. Similarly, the sequences of the transitions from $s_d$ to $s_a$, from $s_i$ to $s_n$, and from $s_q$ to $s_r$ are defined to be single paths. Therefore, it is guaranteed that transitions to $s_a$, $s_n$, or $s_r$ in $RS$ are possible from all states on these paths. As a result, from all states reachable from a state in $RS$, the states in $RS$ can be reached. That is, C6 is satisfied. If a state in $RS$ is selected as the initial state $s_0$, C5 is satisfied. 

Finally, we define $M$ satisfying C1 to C6. By searching for $M$, we can obtain the operation procedure for executing any quantum circuit with at most 56 qubits. Moreover, since the depth-first search does not enter a subspace with no solution, the operation procedure can be obtained in a practical amount of time without backtracking.


\section{Search Algorithm}\label{algorithm}
Even when restricting the transitions of $M$ as described in Section \ref{instance}, several choices arise in the search for transition paths. First, it is necessary to decide which state in $RS$ is to be selected as the initial state $s_0$. The number of shuttling operations can be reduced if the two electrons on which a two-qubit gate is executed are placed close together. This means that the initial placement of the electrons affects the number of shuttling operations. 
Secondly, when executing a single-qubit gate, we can choose whether to evacuate electrons to the right or to the left. Also, when executing a two-qubit gate, we can choose whether to move electron $b_{j1}$ next to $b_{j2}$ or move $b_{j1}$ next to $b_{j2}$. In addition, after the two-qubit gate is executed, if there are a number of nearest seat dots, we can choose which one to move to. These selections can be arbitrarily determined to obtain an operation procedure to execute a given quantum circuit. However, it is better if we can choose as efficient an operation procedure as possible.
Therefore, we apply heuristics to choose efficient transitions, as shown in Algorithm \ref{algo}.
\IncMargin{1em}
\begin{algorithm}
\setstretch{0.9} 
\DontPrintSemicolon

 \SetKwInOut{Input}{input}
 \SetKwInOut{Output}{output}

\Input{$A$, $R$, $QC$}
\Output{$p$}
\BlankLine

$p$, $s_{cur}$ = initial\_placement($A$, $R$, $QC$) \label{a01} \;
$Executed\_ops$ = $\varnothing$ \label{a02} \;
$F$ = front\_layer($QC$, $Executed\_ops$) \label{a03} \;
\While{$F \neq \varnothing$} { \label{a04}
  \For{$f$ in $F$} { \label{a05}
    \uIf{$f.operation$ is g1} { \label{a06}
      $v_{c_j,r_j}$ = pos\_cur($f.electron$) \label{a07} \;
      \uIf{$c_{j} \geq (COL-2)$} { \label{a08}
        $p$, $s_{cur}$ = p.update(g1\_left, $f.electron$)) \label{a09} \;
      }
      \uElseIf{$c_{j} \leq 3$} { \label{a10} 
        $p$, $s_{cur}$ = p.update(g1\_right, $f.electron$)) \label{a11} \;
      }
      \uElseIf{count\_cost(g1\_left, $A$, $s_{cur}$) $>$ count\_cost(g1\_right, $A$, $s_{cur}$)} { \label{a12} 
        $p$, $s_{cur}$ = p.update(g1\_right, $f.electron$)) \label{a13} \;
      }
      \Else{ \label{a14}
        $p$, $s_{cur}$ = p.update(g1\_left, $f.electron$)) \label{a15} \;
      }
    }
    \uElseIf{$f.operation$ is g2} { \label{a16}
      $Pao$ = placement\_after\_ops(f.operation, $A$, $s_{cur}$) \label{a17} \;
      $best\_pao$ = $Pao$[0] \label{a18} \;
      \For{$pao$ in $Pao$} { \label{a19}
        $score$ = eval\_placement($pao$, $QC$, $Executed\_ops$) \label{a20} \;
        \If{$score > best\_pao.score$} { \label{a21}
          $best\_pao$ = $pao$ \label{a22} \;
        }
      }
      $p$, $s_{cur}$ = p.update($best\_pao$, $f.electron1$, $f.electron2$) \; \label{a23}
    }
    \ElseIf{$f.operation$ is m} { \label{a30}
      $p$, $s_{cur}$ = p.update(m, $f.electron$) \; \label{a31}
    }
  }
  $Executed\_ops$ = $Executed\_ops$ $\cup$ $\{ f \}$ \; \label{a24}
  $F$ = front\_layer($QC$, $Executed\_ops$) \; \label{a25}
}
\Return $p$ \label{a26}
\caption{Search Algorithm}\label{algo} 
\end{algorithm}\DecMargin{1em}
The function initial\_placement in line \ref{a01} determines the initial state $s_0$ from the array topology $A$, the structure of the row-shared control gates $R$, and input quantum circuit $QC$. The path $p$ has a list of explored states, i.e., $[s_0]$. The determined state $s_0$ is substituted to the current state $s_{cur}$. The algorithm of initial\_placement is based on the approach of PAQCS \cite{PAQCS}. An electron on which many two-qubit gates are executed with other electrons should be placed in a seat dot that is close to as many electrons as possible. Also, pairs of electrons on which many two-qubit gates are executed should be placed as close together as possible. Moreover, in our SQDA, since the measurement can be conducted only in the rightmost column, the electrons to be measured should be placed as close to the rightmost column as possible.
In line \ref{a03}, the function front\_layer returns the front layer $F$ \cite{sabre}, which is the set of operations that can be executed next. $Executed\_ops$ denotes the set of operations that have been executed. In other words, $F$ is the set of operations that do not have an unexecuted predecessor in the DAG representing $QC$.
When the operation obtained from $F$ is a single-qubit gate (line \ref{a06}), check whether it is possible to secure the evacuation space for other electrons on the right sides of the column $c_j$ where the electron to be operated is located. If not (line \ref{a08}), the other electrons in column $c_{j}$ are moved to column $c_{j}-2$ on the left side by the transitions shown in Fig. \ref{trans1qubit}. The method p.update in line \ref{a09} adds these transitions to the end of $p$ and updates the current state $s_{cur}$. Conversely, if no space can be secured on the left side (line \ref{a10}), the other electrons are moved to the right column $c_{j}+2$ (line \ref{a11}). If the space can be secured on both sides, the direction of the evacuation is determined on the basis of the total distance traveled by the electrons, which is calculated by the function count\_cost (lines \ref{a12}--\ref{a15}).

When the operation obtained from $F$ is a two-qubit gate (line \ref{a16}), we choose which electron is moved to the other side. Moreover, if there is more than one nearest seat dot to which the electron is to be returned, one of them is selected.
In line \ref{a17}, the function placement\_after\_ops extracts possible electron placements after the execution of the two-qubit gate by simulating all those cases. The result is saved in the variable $Pao$. In line \ref{a20}, eval\_placement evaluates the electron placement in terms of the efficiency to execute subsequent two-qubit gates, which is the same as initial\_placement. The most efficient placement is then assigned to $best\_pao$ (line \ref{a22}). The transition from $s_{cur}$ to the state represented by $best\_pao$ is added to the end of $p$, and $s_{cur}$ is updated (line \ref{a23}).

When the operation retrieved from $F$ is a measurement (line \ref{a30}), the transition path according to Figs. \ref{transmeasure} and \ref{returnmeasure} is added to the end of $p$ and $s_{cur}$ is updated (line \ref{a31}). Line \ref{a24} adds the retrieved operation $f$ to the set of executed operations $Executed\_ops$. Front layer $F$ is updated to the latest in line \ref{a25}. If all operations comprising the quantum circuit $QC$ have been executed, $F$ is empty. Then, the loop is broken at line \ref{a04}. After $p$ is returned at line \ref{a26}, the procedure ends.

\section{Experimental Results}\label{exp}
We implemented a quantum compiler (compiler 1) that searches the state transition system $M$ constructed as in Section \ref{instance} with a naive depth-first search, using Qiskit\textregistered \cite{cross2018ibm}. Compiler 1 does not evaluate possible directions to evacuate neighboring electrons by function count\_cost and possible electron placements when returning to a ready state by function eval\_placement in Algorithm \ref{algo}; rather, it adopts the first evacuation direction and electron position it finds. We also implemented a second compiler (compiler 2) that searches $M$ according to Algorithm \ref{algo}. Quantum circuits consisting of the native gates of our SQDA, i.e., $Rx(\theta)$, $Ry(\theta)$, and $(SWAP)^{\alpha}$ gates, are randomly generated and input to these quantum compilers. Then, the compilers generate operation procedures for the input quantum circuit. Compiler 1 outputs the first transition path it finds as the solution. 
The experiment was performed on an Ubuntu 20.04.2 LTS machine equipped with two Intel\textregistered \ Core\texttrademark \ i9-9900K 3.60GHz processors with 8 cores and 62-GB memory. It also has an NVIDIA\textregistered \ GetForce\textregistered RTX 2080 Ti GPU.
\begin{figure*}[htb]
\centering
\scalebox{0.45}{\includegraphics{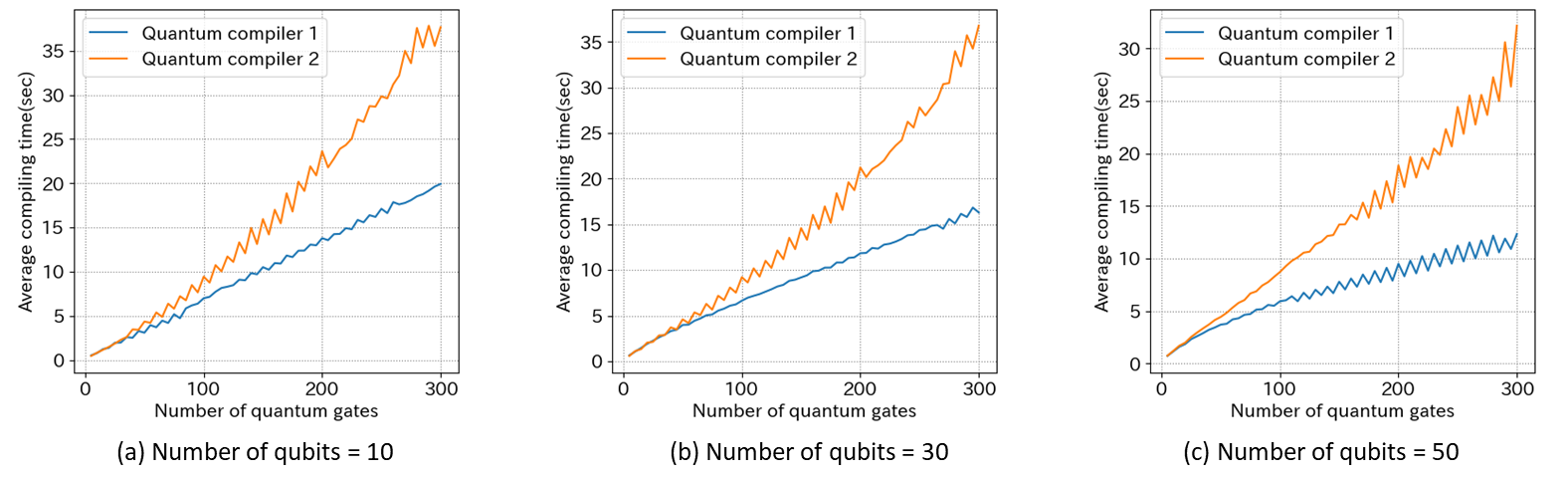}}
\caption{Average compiling time with increasing number of quantum gates}
\label{exp_gatetime}
\end{figure*}
\begin{figure*}[htb]
\centering
\scalebox{0.45}{\includegraphics{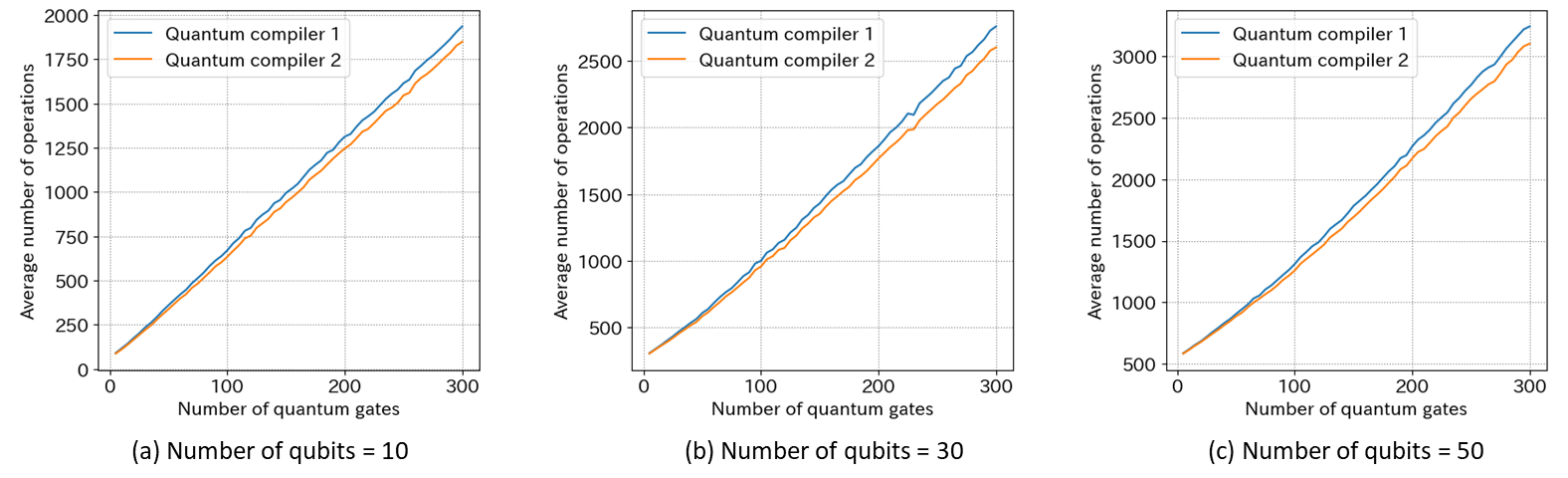}}
\caption{Average number of operations with increasing number of quantum gates}
\label{exp_gateoperation}
\end{figure*}

Figure \ref{exp_gatetime} shows the average compiling time of 100 quantum circuits when the number of quantum gates in the quantum circuit is increased from 1 to 300 for 10, 30, and 50 qubits, respectively. 
It is confirmed here that the operation procedures for our SQDA can be generated in a practical amount of time, which indicates that there is a solution in $M$ constructed by the proposed method and that we do not spend a huge amount of time searching a subspace with no solution. We also find that the compiling time increases with the number of quantum gates but is not significantly affected by the number of qubits. This is because the compilers refer to each quantum gate one by one from the quantum circuit and generate the corresponding operation procedures. 
Comparing compilers 1 and 2 in Fig. \ref{exp_gatetime}, we can see that the compile time of 1 is shorter than that of 2. This is due to the fact that compiler 2 considers possible evacuation directions and electron placements as shown in Algorithm \ref{algo}. Figure \ref{exp_gateoperation} shows the average number of shuttling, quantum gate, and measurement operations output by the compilers in the same experiment as in Fig. \ref{exp_gatetime}. This indicate that compiler 2 with Algorithm \ref{algo} can generate a more efficient operation procedure than compiler 1 with the naive depth-first search. However, the effect is not significant. Improvements to generate more efficient operation procedures are discussed in Section \ref{improvement}.


We also implemented a program to check the constraints and requirements represented by formulae \ref{f1qubit} to \ref{freq2} and were able to confirm that the operation procedures for our SQDA generated by compilers 1 and 2 satisfy these constraints and requirements. This means that the generated procedures can execute the target quantum circuits on the actual SQDA without crosstalk or collisions of the electrons.


\section{Discussion}\label{discuss}
\subsection{Improvements of State Transition System}\label{improvement}
The proposed method constructs state transitions to conduct a single-qubit gate, two-qubit gate, and the measurement independently, and restricts the other state transitions. This guarantees that any operation can be executed in any order. However, because of the restriction on $M$, it is not possible to generate an operation procedure that simultaneously executes an operation on multiple electrons at a time. For example, in the quantum circuit shown in Fig. \ref{dag}, the same $Rx(\theta)$ gate is first executed on qubit $q_0$ and then on $q_1$. Assume the electrons corresponding to $q_0$ and $q_1$ are $b_0$ and $b_1$, respectively. If $b_0$ and $b_1$ are initially placed in the same column, the $Rx(\theta)$ gate can be efficiently executed on $b_0$ and $b_1$ simultaneously by using the column-shared control gates.
However, with our current $M$, a procedure that executes the $Rx(\theta)$ gate on $b_0$ followed by the same gate on $b_1$ is extracted. This increases the number of operations to achieve the required quantum computation. 
Similarly, some control gates can be operated in parallel, which also enables us to execute multiple operations simultaneously. Therefore, to generate a more efficient operation procedure, it is necessary to relax the constraints of $M$, i.e., to allow other transitions in $M$ so that these simultaneous operations can be extracted.

\begin{figure}[htb]
\centering
\scalebox{0.28}{\includegraphics{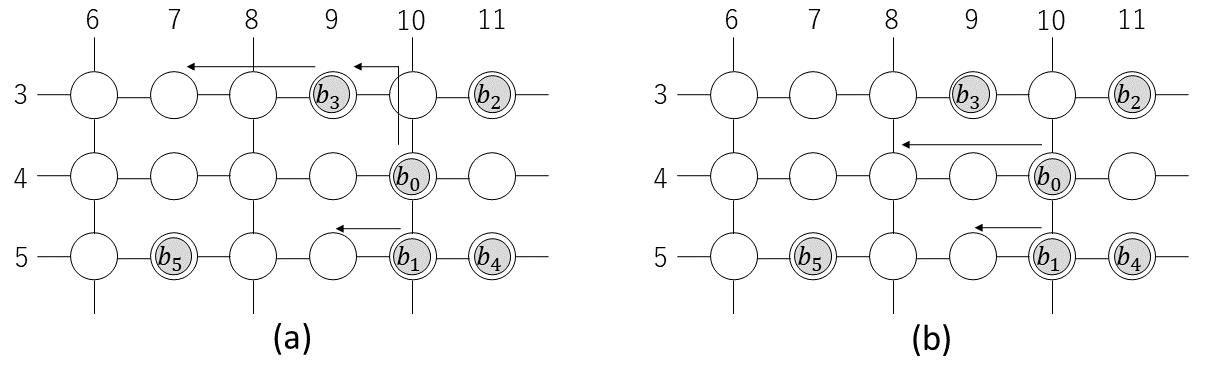}}
\caption{More efficient movements of electrons}
\label{billiards}
\end{figure}

Allowing other transitions of $M$ means we can obtain a more efficient operating procedure.
In the transition discussed in Section \ref{sec-trans1qubit}, the electrons move to the nearest empty seat dots after the two-qubit gate is executed. 
However, to efficiently execute the subsequent two-qubit gates, moving to a non-empty seat dot should also be considered. For example, Fig. \ref{billiards} shows the state after the execution of a two-qubit gate on $b_0$ and $b_1$. After $b_1$ goes back to quantum dot $v_{5,9}$, the nearest empty seat dot from $b_0$ is $v_{3,7}$. We assume that the subsequent quantum gate operations include a two-qubit gate of $b_0$ and $b_2$ and do not include quantum gate operations using $b_3$. In that case, instead of moving $b_0$ to $v_{3,7}$, it is more efficient to first move $b_3$ to $v_{3,7}$ and then move $b_0$ to $v_{3,9}$, as shown Fig. \ref{billiards}(a). By pushing $b_3$ out in this way, $b_0$ and $b_2$, on which the subsequent two-qubit gate will be executed, can be placed close together. In fact, if $b_0$ is moved to $v_{3,7}$, four shuttling operations are conducted. In addition, five shuttling operations are necessary for the execution of the subsequent two-qubit gate on $b_0$ and $b_2$. In contrast, when moving $b_0$ to $v_{3,9}$, four and three shuttling operations are required, respectively. This means that the total number of operations can be reduced by pushing electrons.

If an electron utilized for a two-qubit gate is also used for the immediately following two-qubit gate, it can move directly to the next two-qubit gate position without returning to a seat dot. For example, in Fig. \ref{billiards}(b), it is assumed that the two-qubit gate on $b_0$ and $b_5$ will be executed immediately after the two-qubit gate on $b_0$ and $b_1$. In that case, it is more efficient to move $b_0$ directly to $v_{4,8}$ than to move once back to $v_{3,9}$ or $v_{3,7}$. Thus, we should consider direct transitions from a state in $OS$ to the next state in $OS$ without going back to a state in $RS$.



\subsection{Effectiveness of Shuttling for Crosstalk Avoidance}
\begin{figure}[htb]
\centering
\scalebox{0.28}{\includegraphics{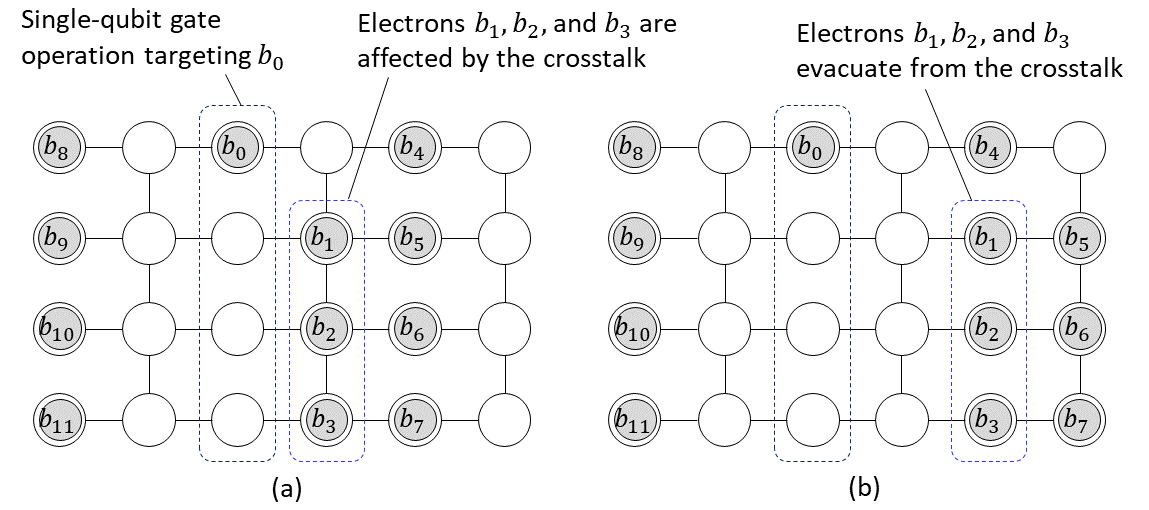}}
\caption{Single-qubit gate operation in the cases of with-crosstalk and w/o-crosstalk}
\label{with_and_without_crosstalk}
\end{figure}

As discussed in Section \ref{sec-trans1qubit}, when executing a single-qubit gate, crosstalk can be avoided if we evacuate neighboring electrons by shuttling. However, the fidelity of the quantum state is also reduced by the shuttling operation. Therefore, the reduction in fidelity due to the shuttling operation should be smaller than the reduction in fidelity due to crosstalk. 
In this section, we evaluate the fidelity reduction in two cases: one in which crosstalk is not avoided (with-crosstalk) and one in which it is avoided by shuttling (w/o-crosstalk). Examples of a single-qubit gate operation for the cases with-crosstalk and w/o-crosstalk are shown in Figs. \ref{with_and_without_crosstalk}(a) and (b), respectively. In the case of with-crosstalk, the electrons $b_1$, $b_2$, and $b_3$, which were in the same row as the target electron $b_0$, are left in the adjacent row and are affected by the cross-talk of the single-qubit gate. In contrast, in the case of w/o-crosstalk, $b_1$, $b_2$, and $b_3$ are evacuated to two columns away from the column of $b_j$, so they are not affected by the crosstalk.

We assume that only crosstalk and shuttling operations cause the reduction in fidelity, and do not consider other factors (e.g., quantum gate and measurement operations).
In the case of w/o-crosstalk, we estimate the number of shuttling operations in the procedure extracted by the algorithm shown in Section \ref{instance}. Let $n$ and $m$ be the number of qubits and quantum gates constituting the target quantum circuit, respectively.
In our 16 $\times$ 8 SQDA, the electrons are placed in every row (except the bus row) and in every two rows, as shown in Fig. \ref{electronposition}. The lengths of the array sides are $O((1/2)\sqrt{2n})$ and $O(2 \sqrt{2n})$, respectively. The number of electrons in each row and each column are $O(\sqrt{n})$. As defined in Fig. \ref{trans1qubit}, when executing a single-qubit gate on electron $b_j$ in $v_{r_j,c_j}$, all electrons in column $c_{j-2}$ are moved to column $c_{j-3}$, and then all electrons in column $c_j$ (except $b_j$) electrons are moved to column $c_{j-2}$. Therefore, the number of shuttling operations required before executing the single-qubit gate is $O(3 \sqrt{n})$. After the single-qubit gate is executed, the reverse procedure is executed to return to the original state in $RS$. Therefore, the number of shuttling operations after the single-qubit gate is also $O(3 \sqrt{n})$. The same is true when evacuating electrons to the right side. The number of shuttling operations required for single-qubit gates is expressed as $O(6m \sqrt{n})$.

In the case of a two-qubit gate, as defined in Fig. \ref{trans2qubit}, electrons $b_{j1}$ and $b_{j2}$ are moved through the aisle column and bus row to be adjacent. The number of shuttling operations depends on the distance between $b_{j1}$ and $b_{j2}$.
For the sake of simplicity, we assume that the number of shuttling operations is its median value. The maximum is the sum of the lengths of the array sides, that is, $(1/2)\sqrt{2n} + 2 \sqrt{2n}$, and the minimum is 3. Therefore, it is assumed that $O((1/4)\sqrt{2n} + \sqrt{2n}) = O((5/4)\sqrt{2n})$.
Although $b_{j1}$ and $b_{j2}$ are returned to the nearest seat dot after the two-qubit gate is executed (as defined in Fig. \ref{return2qubit}), we assume that they are returned to the original seat dot. Therefore, the number of shuttling operations for two-qubit gates is expressed as $O((5/2)m \sqrt{2n})$.

In the case of a measurement, if the bus row is placed approximately in the center of the array, the average number of vertical shuttling operations to the bus row is $O((1/8) \sqrt{2n})$. The average number of horizontal shuttling operations is $O(\sqrt{2n})$. In the generated quantum circuits, all qubits are measured once at the end. Therefore, the number of shuttling operations for measurements is estimated as $O((9/8)n \sqrt{2n})$. 
Finally, the number of shuttling operations in the case of w/o-crosstalk is $O(6m \sqrt{n} + (5/4)m \sqrt{2n} + (9/8)n \sqrt{2n})$

Next, in the case of with-crosstalk, we estimate the number of shuttling operations and crosstalk occurrences. When executing a single-qubit gate, electrons in column $c_j$ (other than $b_j$) do not have to evacuate to column $c_{j-2}$, where they are not affected by crosstalk. However, if they are left on $c_j$, the single-qubit gate will be executed on them together with $b_j$. Therefore, they are moved to column $c_{j-1}$. In this case, the number of shuttling operations is reduced from $O(6m \sqrt{n})$ to $O(2m \sqrt{n})$. All electrons moving from column $c_j$ to column $c_{j-1}$ are affected by crosstalk. The number of crosstalk occurrences is $O(m \sqrt{n})$.

In the case of a two-qubit gate, the electrons in the same row as the target electrons are located in the seat dots that are not connected to the neighboring dots by channels. Therefore, they are not affected by the two-qubit gate operation, and the number of shuttling operations is the same as in the case of w/o-crosstalk. As for a measurement, the operation procedure is the same as in the case of w/o-crosstalk. Finally, the number of shuttling operations in the case of with-crosstalk is $O(2m \sqrt{n} + (5/4)m \sqrt{2n} + (9/8)n \sqrt{2n})$, and the number of crosstalk occurrences is $O(m \sqrt{n})$.

Denote the fidelity of a shuttling operation as $f_{sh}$ and the fidelity of crosstalk as $f_{ct}$. The fidelity reduction due to shuttling can be estimated by multiplying the fidelity of one shuttling operation by the number of shuttling operations. The same is true for crosstalk.
The fidelity in the case of w/o-crosstalk is estimated as $O(f_{sh}^{6m \sqrt{n} + (5/4)m \sqrt{2n} + (9/8)n \sqrt{2n}})$. Similarly, the fidelity in the case of with-crosstalk is $O(f_{sh}^{2m \sqrt{n} + (5/4)m \sqrt{2n} + (9/8)n \sqrt{2n}} \cdot f_{ct}^{m \sqrt{n}})$. Let $f_{sh}^{2m \sqrt{n} + (5/4)m \sqrt{2n} + (9/8)n \sqrt{2n}}$ be $a$; then, we can express them as $O(a \cdot f_{sh}^{4m \sqrt{n}})$ and $O(a \cdot f_{ct}^{m \sqrt{n}})$, respectively. Thus, if $f_{sh}^{4} > f_{ct}$ holds, the fidelity of w/o-crosstalk will be higher than that of with-crosstalk.

\begin{figure*}[htb]
\centering
\scalebox{0.45}{\includegraphics{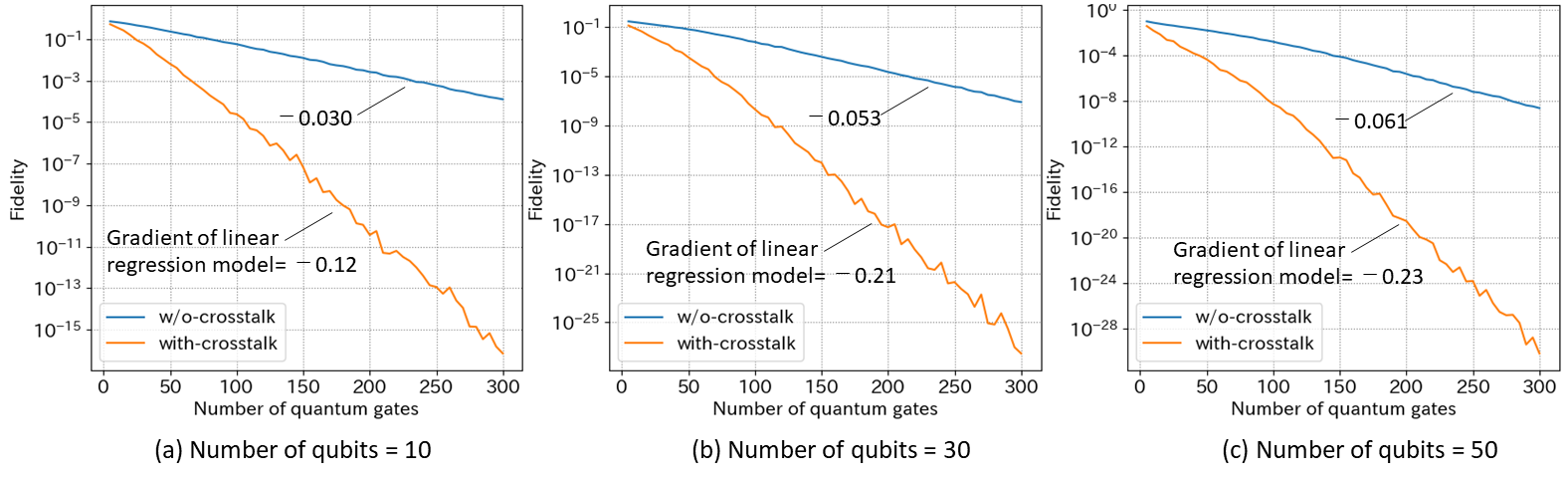}}
\caption{Fidelity with increasing number of quantum gates}
\label{fidelity_gate}
\end{figure*}

We experimentally calculated fidelities for the quantum circuits used in Section \ref{exp} in the cases of w/o-crosstalk and with-crosstalk by compiler 2.
Figure \ref{fidelity_gate} shows the fidelities when the number of qubits of the quantum circuits is 10, 30, and 50, respectively, assuming $f_{sh} = 0.996$ \cite{Noiri2022} and $f_{ct} = 0.905$ \cite{philips2022universal}\footnote{In their previous study, the crosstalk of the Rabi oscillation causes oscillations in the neighboring qubits, whose magnitude is at most 0.2. The fidelity of the crosstalk is then assumed to be $cos^{2}((0.2 * \theta)/2)$, where $ \theta $ is the rotation angle of a single-qubit gate. Assume that $ \theta $ is $ \pi $, where the effect of the crosstalk is maximized.}. The horizontal axes represent the number of quantum gates. 
When $f_{sh} = 0.996$ and $f_{ct} = 0.905$, $f_{sh}^{4} > f_{ct}$ is held; thus, higher fidelity is expected in the w/o-crosstalk cases. In fact, we confirmed that the fidelities of the w/o-crosstalk cases are higher than those of the with-crosstalk cases.

The fidelity in the case of w/o-crosstalk is estimated as $O((0.996)^{6m \sqrt{n} + (5/4)m \sqrt{2n} + (9/8)n \sqrt{2n}})$. Similarly, the fidelity in the case of with-crosstalk is $O((0.996)^{2m \sqrt{n} + (5/4)m \sqrt{2n} + (9/8)n \sqrt{2n}} \cdot (0.905)^{m \sqrt{n}})$. The logarithms for these orders are as follows.
\begin{align}
& O(log((0.996)^{6m \sqrt{n} + \frac{5}{4}m \sqrt{2n} + \frac{9}{8}n \sqrt{2n}})) \nonumber \\
& = O((6m \sqrt{n} + \frac{5}{4}m \sqrt{2n} + \frac{9}{8}n \sqrt{2n}) \cdot (-0.00174)) \label{wocross} \\
& \simeq O(8m \sqrt{n} \cdot (-0.00174)) \label{wocross_aprx}
\end{align}
\begin{align}
& O(log((0.996)^{2m \sqrt{n} + \frac{5}{4}m \sqrt{2n} + \frac{9}{8}n \sqrt{2n}} \cdot (0.905)^{m \sqrt{n}})) \nonumber \\
& = O((2m \sqrt{n} + \frac{5}{4}m \sqrt{2n} + \frac{9}{8}n \sqrt{2n}) \cdot (-0.00174) \nonumber \\
& \quad + m \sqrt{n} \cdot (-0.0434)) \label{withcross} \\
& \simeq O(4m \sqrt{n} \cdot (-0.00174) + m \sqrt{n} \cdot 25(-0.00174)) \nonumber \\
& = O(29m \sqrt{n} \cdot (-0.00174)) \label{withcross_aprx}
\end{align}
From formulae \ref{wocross_aprx} and \ref{withcross_aprx}, we estimate that the gradient of the fidelity for $m$ in the case of with-crosstalk is almost 3.6 times that of w/o-crosstalk. In fact, the gradients in the case of with-crosstalk are 4.1, 3.9, and 3.7 times those of w/o-crosstalk in Fig. \ref{fidelity_gate}, respectively. These experimental results are consistent with the theoretical estimation. 
When focusing on the number of qubits $n$, formulae \ref{wocross} and \ref{withcross} are approximated as $O( (1.6n + 7.8m) \sqrt{n} \cdot (-0.00174) )$ and $O( (1.6n + 28.8m) \sqrt{n} \cdot (-0.00174) )$, respectively.They are of higher order for $n$ than for $m$, but the coefficient of $m$ is larger than that of $n$. Therefore, when $n$ is small, $m$ contributes more strongly to the decrease in fidelity. However, as $n$ becomes larger, it has a greater impact to the decrease in fidelity than $m$.

\begin{figure*}[htb]
\centering
\scalebox{0.45}{\includegraphics{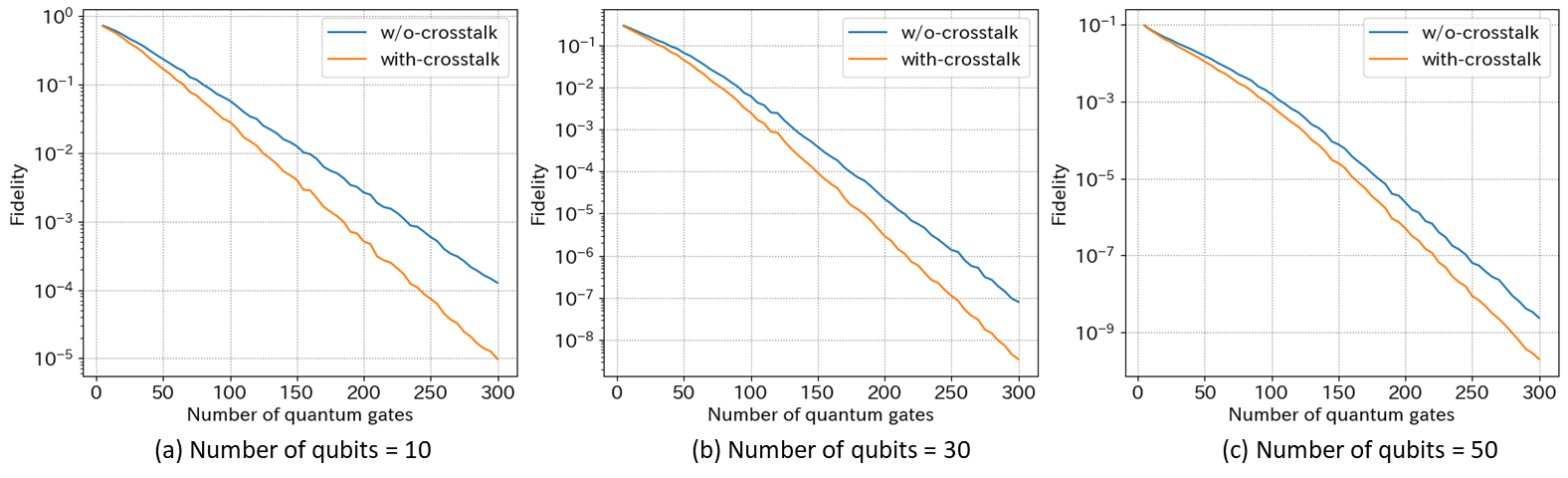}}
\caption{Fidelity with increasing number of quantum gates when $f_{sh}^4 = f_{ct}$}
\label{fidelity_gate2}
\end{figure*}

Figure \ref{fidelity_gate2} shows the results when $f_{sh} = 0.996$ and $f_{ct} = 0.984$, so that $f_{sh}^{4} = f_{ct}$.
The fidelities of with-crosstalk and w/o-crosstalk are almost equal, as expected. From formulae \ref{wocross} and \ref{withcross}, the errors in the number of shuttling operations and crosstalk increase exponentially with the increase in the number of quantum gates $m$. Therefore, in Fig. \ref{fidelity_gate2}, the difference between with-crosstalk and w/o-crosstalk increases with the number of quantum gates $m$ (note that the fidelity is plotted on a logarithmic scale). This confirms that if $f_{sh}^{4} > f_{ct}$ is satisfied, avoiding crosstalk will result in higher fidelity.

\begin{figure*}[htb]
\centering
\scalebox{0.45}{\includegraphics{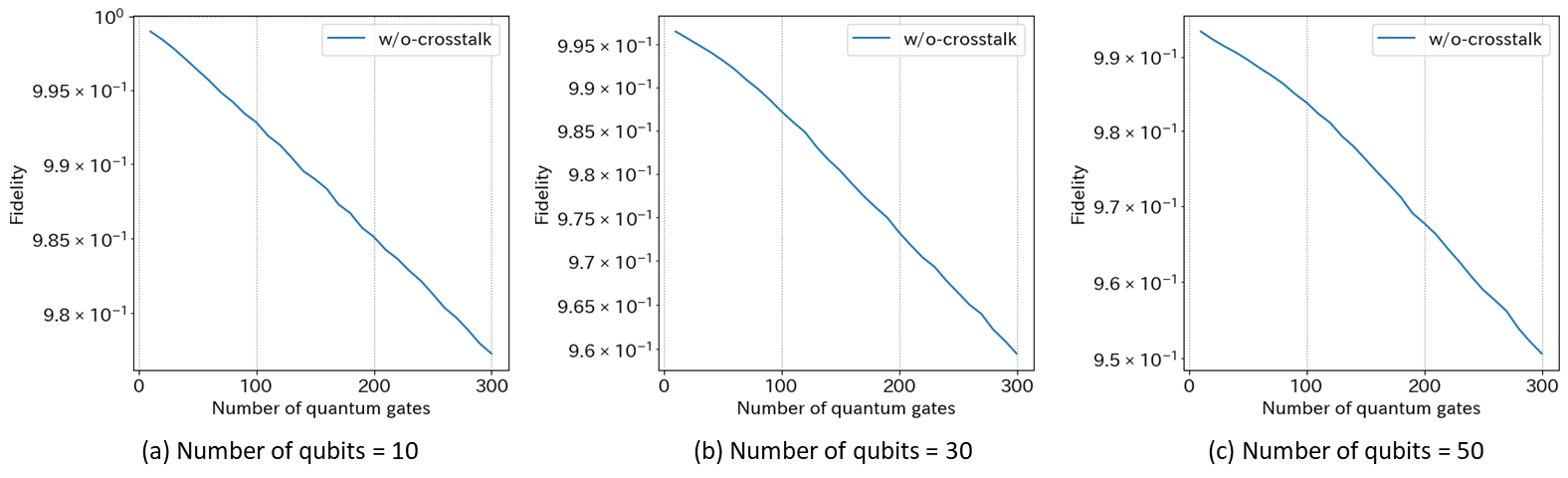}}
\caption{Fidelity with increasing number of quantum gates when $f_{sh}^4 = 0.99999$}
\label{fidelity_gate3}
\end{figure*}

Figure \ref{fidelity_gate3} shows the fidelities of w/o-crosstalk when $f_{sh} = 0.99999$ and $f_{ct} = 0.905$. In this case, even when the number of qubits is 50 and the number of quantum gates reaches 300, the fidelity is maintained over 0.95. This suggests that the improvement in shuttling fidelity is necessary for future practical use. Moreover, as discussed in Section \ref{improvement}, the proposed method should also be improved to reduce the number of shuttling operations.



\section{Related Work}\label{relwork}
We have explaind how to generate operation procedures to execute an arbitrary quantum circuit in a strictly constrained silicon quantum dot array. As mentioned in Section \ref{problem}, there is a similar problem called qubit mapping problem \cite{zhu2020dynamic}\cite{zhu2021iterated}\cite{sabre} in superconducting quantum computing, which minimizes the number of SWAP gates to be inserted to satisfy the NN-constraint.
In the array of a superconducting quantum computer, the target qubits must be adjacent to each other to execute a two-qubit gate. When the target qubits are not adjacent, the contents of the qubits are transferred by inserting SWAP gates. The number of SWAP gates inserted to satisfy the NN-constraint is called the nearest neighbor cost (NN-cost) \cite{Robert2016}\cite{sabre}. The route to transfer the contents of the qubits should be selected so that the NN-cost is as small as possible.
If the target qubits are located far apart, a lot of SWAP gates are inserted, so it is desirable to initially position the target qubits of the two-qubit gate as close together as possible. Moreover, transferring the content of the target qubit by inserting SWAP gates also changes the contents of other qubits on the route. If those other qubits are used in a subsequent quantum gate, it is necessary to take the subsequent quantum gates into consideration when determining the movement route of the qubit. Various methods for optimizing the initial placement and movement routes of qubits have been proposed for 1-dimensional arrays \cite{wille2010synthesis}\cite{chakrabarti2011linear}\cite{saeedi2011synthesis}\cite{Hirata2011}\cite{gel2012}\cite{shafaei2013optimization}\cite{wille2014optimal}\cite{rahman2015synthesis} and 2-dimensional arrays \cite{Alireza2014}\cite{determiningminimal}\cite{PAQCS}\cite{Robert2016}\cite{efficientmapping}\cite{bhattacharjee2017depthoptimal}\cite{PAQCS2}\cite{Bhattacharjee2018ANA}\cite{gel2019}\cite{sabre}\cite{tiket}\cite{matsuo2023sat}\cite{li2023quantum}.
These studies tackled the optimization problem of minimizing the NN-cost, i.e., the number of SWAP gates to be inserted. Most of them presented heuristics to find an efficient operation procedure from among a huge number of candidates. Only executing the target quantum circuit by inserting SWAP gates is not difficult as long as its efficiency is not a concern.
In contrast, in our SQDA, the procedure to execute the target quantum circuit is non-trivial due to the complex constraints of the array (as formally described in Section \ref{constraint}). Therefore, we presented a method that enables us to obtain the procedures to execute arbitrary target quantum circuits by restricting the initial placement and movements of the electrons. We also introduced the heuristics of Algorithm \ref{algo} to obtain as efficient procedures as possible. The previous studies tackling the qubit mapping problem correspond to Algorithm \ref{algo} in this paper. In fact, function initial\_placement in Algorithm \ref{algo} refers to an approach proposed in one of the previous studies \cite{PAQCS}.

If some two-qubit gates consisting of a quantum circuit are commutative, the execution order of them can be changed. Leveraging this property, a method was proposed to change the execution order of two-qubit gates so that the NN-cost is as small as possible \cite{gel2012}\cite{gel2019}. Since our native two-qubit gate, $(SWAP)^{\alpha}$, is not commutative, this method cannot be applied to our algorithm directly. However, the idea of transforming the quantum circuit to reduce NN-cost is useful for us. Ferrari et al. proposed transforming VQE ansatz, which consists of a CNOT cascade, into an equivalent quantum circuit consisting of CNOT gates on linearly adjacent qubits \cite{Ferrari2021}. This allows all CNOT gates to be executed without inserting SWAP gates when the qubits are placed linearly adjacent to each other. Then, the NN-cost is greatly reduced.
Similarly, synthesis methods have also been proposed, in which the required quantum computation is achieved with as few quantum gates as possible \cite{ExactSynthesis}\cite{numericalanalysis}. If the number of quantum gates can be reduced bu utilizing these techniques, the length of the operation procedure can also be shortened.

To improve the fidelity of the calculated quantum state, taking into account the error rate of physical qubits and their links when solving the qubit mapping problem has been proposed \cite{NoiseAdaptive}\cite{NotAllQubits}\cite{MachineLearningBased}. To make the shuttling operation procedure more efficient in our SQDA, the device error characteristics of quantum gate operations and measurements should be considered.
Moreover, to reduce crosstalk, optimizing the scheduling of quantum gate operations and the location of the gate execution on the array has been proposed \cite{softcrosstalk1}\cite{softcrosstalk2}. In this work, we have defined crosstalk avoidance as a constraint and generate an operation procedure that satisfies the constraint. This is similar to these previous studies in that crosstalk is reduced by compiler software.

In trapped-ion and trapped-atom quantum computers, the qubits can be moved by shuttling the ions and atoms that form the building blocks of the qubits, as in silicon quantum computers. Therefore, they are moved to satisfy the NN-constraint to execute a two-qubit gate. 
In trapped-ion quantum computers, two ions must be stored in the same trap, which is achieved by shuttling the target ions. There have been several studies on improving the efficiency of shuttling procedures for 1-dimensional \cite{Muzzle}\cite{Durandau2023automatedgeneration} and 2-dimensional architectures \cite{EfficientQubitRouting}. Webber et al. proposed a 2D array structure to allow any ion to move to any trap by shuttling and defined the direction properties of ion passage on the array \cite{EfficientQubitRouting}. This allows for a two-qubit gate to be executed on any pair of ions. Their aim was to establish ways of executing arbitrary quantum circuits, which is similar to our aim. Restricting the direction of ion passage in their study corresponds to restricting the transitions of $M$ in this paper. However, our SQDA has more complex restrictions compared to trapped-ion arrays, such as the fact that electrons in the same row move together. Therefore, we restrict the movement of electrons more strictly: for example, electrons can basically only pass through the bus row and aisle column.
In trapped-atom quantum computers, to execute a two-qubit gate, the target atoms are moved by an optical tweezer so that they are adjacent to each other. Prior studies have attempted to optimize the movement of the target atoms \cite{Brandhofer2021Optimal}\cite{qubitmappingatom2022} by using a Satisfiability Modulo Theories (SMT) model that incorporates physical constraints for the movement by the optical tweezer. The model is then input to an SMT solver \cite{smtsolver} and the optimal procedure is obtained as a solution. By describing the constraints of our SQDA in the SMT model, it may be possible to derive an optimal operation procedure. However, since the SMT solver is not scalable, this method is difficult to apply to large-scale quantum circuits.

In silicon quantum computers, there are many previous studies on the implementation of shuttling \cite{Spiderweb}\cite{Mills2019}\cite{BlueprintofaScalableSpin}\cite{Acrossbarnetwork}\cite{Simulatedcoherentelectron}\cite{ShuttlinganElectronSpin}\cite{Fujita2017}\cite{Nakajima2018}\cite{Yoneda2021}\cite{Noiri2022}, which is the premise of this paper. However, as far as we know, there have been no studies on the pathways for transferring electrons by shuttling, or on their optimization. We show for the first time how to compile an arbitrary quantum circuit and generate operation procedures for silicon quantum computers.




\section{Conclusion}\label{conclusion}
We proposed a method to generate an operation procedure to execute arbitrary quantum circuits in our SQDA, which has complex constraints. We first defined a state transition system $M$ as a formal model of the SQDA. Then, using this model, we formally described the constraints and requirements that the operation procedure should satisfy. Second, we discussed the problem that occurs when searching for operation procedures for the target quantum circuit, which is that there may be no solution in the entire search space or a particular subspace. This results in a search space having to be exhaustively explored, and the operation procedure cannot be obtained in a practical amount of time. To solve this problem, we introduced six conditions that should be satisfied in $M$ and demonstrated the concrete $M$ for our SQDA satisfying those conditions. We also implemented the concrete $M$ and its search algorithm as quantum compilers. Experiments with these compilers demonstrated that operation procedures for our SQDA can be discovered in a practical amount of time by searching $M$. 

Future work will include implementing the improvements of $M$ described in Section \ref{improvement}. The restriction of transitions in $M$ should be relaxed to generate more efficient operation procedures. 
The core idea of electron movements presented in this paper consists of bus row, aisle column and seat dot, but there may be other ideas to extract more efficient operation procedures. In particular, if we restrict the target quantum circuit to a particular algorithm, we may be able to construct a more efficient $M$ specific to that algorithm. For example, for quantum circuits implementing the Quantum Approximate Optimazation Algorithm (QAOA), there is a certain pattern in the order of the two-qubit gates to be executed. By focusing on that pattern, it may be possible to develop an $M$ that is more efficient for the quantum circuit of QAOA, but which may not support other quantum circuits. This approach should be effective over the next few years because only a limited number of algorithms are of practical use in the Noisy Intermediate-Scale Quantum (NISQ) era \cite{Preskill2018quantumcomputingin}.

\section*{Acknowledgment}
This work was supported by JST Moonshot R\&D Grant Number JPMJMS2065.

\bibliographystyle{IEEEtran}
\bibliography{shuttling}

\section*{Appendix}
\subsection{Definitions of predicates}\label{predicatedef}

\begin{align}
& SameCol_i(b_j, b_k) \stackrel{\mathrm{def}}{=} \nonumber \\
& v_{r_j,c_j} = pos_i(b_j) \land v_{r_k,c_k} = pos_i(b_k) \land c_j = c_k
\end{align}
\begin{align}
& SameRow_i(b_j, b_k) \stackrel{\mathrm{def}}{=} \nonumber \\
& v_{r_j,c_j} = pos_i(b_j) \land v_{r_k,c_k} = pos_i(b_k) \land r_j = r_k
\end{align}
\begin{align}
& AdjHor_i(b_j, b_k) \stackrel{\mathrm{def}}{=} \nonumber \\
& v_{r_{j},c_{j}} = pos_i(b_{j}) \land v_{r_{k},c_{k}} = pos_i(b_{k}) \land r_{j} = r_{k} \nonumber \\
& \land |c_{j} - c_{k}| = 1 \land (v_{r_{j},c_{j}}, v_{r_{k},c_{k}}) \in E
\end{align}
\begin{align}
& AdjVer_i(b_{j}, b_{k}) \stackrel{\mathrm{def}}{=} \nonumber \\
& v_{r_{j},c_{j}} = pos_i(b_{j}) \land v_{r_{k},c_{k}} = pos_i(b_{k}) \land c_{j} = c_{k} \nonumber \\
& \land |r_{j} - r_{k}| = 1 \land (v_{r_{j},c_{j}}, v_{r_{k},c_{k}}) \in E \nonumber \\
& \land R(v_{r_{j},c_{j}}) \land R(v_{r_{k},c_{k}})
\end{align}

\begin{align}
& BC(v_{r_{k1},c_{k1}}, v_{r_{k2},c_{k2}}) \stackrel{\mathrm{def}}{=} \nonumber \\
& \left( R(v_{r_{k1},c_{k1}}) \land \lnot R(v_{r_{k2},c_{k2}}) \land v_{r_{k2},c_{k2}} \notin P_i \right) \nonumber \\
& \lor \left( \lnot R(v_{r_{k1},c_{k1}}) \land R(v_{r_{k2},c_{k2}}) \land v_{r_{k2},c_{k2}} \notin P_i \right) \label{fbccondition}
\end{align}

\begin{align}
Reachable(s_a, s_b) \stackrel{\mathrm{def}}{=} \exists p = [ t_{1}, ..., t_{i}, ..., t_{n} ] \cdot
 s_a = s_b \nonumber \\ 
\lor \left( t_{1} = (s_a, l_{a+1}, s_{a+1}) \land t_{n} = (s_{b-1}, l_{b}, s_{b}) \right) 
\end{align}

\subsection{Exception Case of Transition for Measurements}\label{transmeasure-exception}
\begin{figure}[htb]
\centering
\scalebox{0.4}{\includegraphics{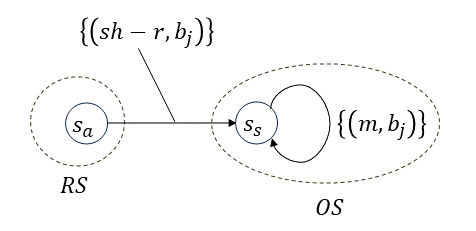}}
\caption{Transitions to execute a measurement in the exception case}
\label{transmeasure_exception}
\end{figure}

If the target electron to be measured is in column 15, it can be moved to column 16 directly by shuttling to the right. The transition in such an exception case is shown in Fig. \ref{transmeasure_exception}.
The transition to return to a state in $RS$ is the same as the one shown in Fig. \ref{returnmeasure}.


\end{document}